\documentclass{aa}

\usepackage{txfonts}

\usepackage[english]{babel}

\usepackage{graphicx}

\usepackage{latexsym}

\begin{document}

\title{Prospects for population synthesis in the H band: \\
NeMo grids of stellar atmospheres compared to observations}

\author{J. Fr\'emaux\inst{1} \and F. Kupka\inst{2} \and C. Boisson\inst{1} \and M. Joly\inst{1} \and V. Tsymbal\inst{3,}\inst{4}}

\institute{LUTH, UMR 8102 du CNRS, associ\'e \`a l'Universit\'e Denis Diderot,
           Observatoire de Paris, Section de Meudon, 92195 Meudon Cedex, France
	   \and Max-Planck-Institute for Astrophysics, Karl-Schwarzschild Str.~1,
	   85741 Garching, Germany
	   \and Tavrian National University, Yaltinskaya~4, 330000 Simferopol,
	    Crimea, Ukraine
	   \and
	    Institute for Astronomy, University of Vienna, T\"urkenschanzstra{\ss}e 17, A-1180 Vienna, Austria}
\offprints{J. Fr\'emaux, \email{julien.fremaux@obspm.fr}}

\date{Received 27 June 2005 / Accepted 22 October 2005}

 
  \abstract
  {For applications in population synthesis, libraries of theoretical stellar spectra are often considered an alternative to template libraries of observed spectra, because they allow a complete sampling of stellar parameters. Most attention in published theoretical spectral libraries has been devoted to the visual wavelength range.}
  {We present a detailed comparison of theoretical spectra in the range 1.57-1.67$\mu$m, for spectral types from A to early M and for giants and dwarf stars, with observed stellar spectra at resolutions around 3000, which would be sufficient to disentangle the different groups of late type stars.}
  {We have selected the NeMo grids of stellar atmospheres to perform such a comparison.}
  {We first demonstrate that after combining atomic and molecular line lists, it is possible to match observed spectral flux distributions with theoretical ones very well for almost the entire parameter range covered by the NeMo grids at moderate resolution in the visual range. In the infrared range, although the overall shape of the observed flux distributions is still matched reasonably well, the individual spectral features are reproduced by the theoretical spectra only for stars earlier than mid F type. For later spectral types the differences increase and theoretical spectra of K type stars have systematically weaker line features than those found in observations. These discrepancies are traced back to stem primarily from incomplete data on neutral atomic lines, although some of them are also related to molecules.}
  {Libraries of theoretical spectra for A to early M type stars can be successfully used in the visual regions for population synthesis but their application in the infrared is restricted to early and intermediate type stars. Improving atomic data in the near infrared is a key element in making the construction of reliable libraries of stellar spectra in the infrared feasible.}


\titlerunning{Prospects for population synthesis in the H band}

\authorrunning{J. Fr\'emaux et al.}

\maketitle

\section{Introduction}

Due to the rapidly increasing spectral resolution of galaxy surveys
(e.g.\ Sloan Digital Sky Survey), modelling any galaxy with spectral
synthesis technique requires a high spectral resolution stellar
library.

Over the last years much progress has been made in synthesis models 
(e.g.\ Pelat \cite{Pel97}; Leitherer et al.\ \cite{Lei99};  Moultaka \& Pelat \cite{Mou00}; Bruzual \& Charlot \cite{Bru03};
 Le Borgne et al.\ \cite{LeB04}; 
Cid Fernandes et al.\ \cite{Cid05}). Moderate to high spectral resolution observations
of stars in order to construct reliable template libraries have also been performed
in the visible domain (e.g.\ STELIB of Le Borgne et al.\ \cite{LeB03}; UVES of
Bagnulo et al.\ \cite{Bag03}; CoudeFed of Valdes et al.\ \cite{Val04}) and in the
IR (e.g.\ Dallier et al.\ \cite{Dal96}; Meyer et al.\ \cite{Mey98}; Ivanov et al.\ \cite{Iva04}). 
However, the major limitation of all these libraries
is the sampling of stellar parameters such as metallicity.

One way to avoid such a difficulty is to build libraries of
theoretical stellar spectra in order to choose the desired physical
parameters.  In this sense, some extensive libraries of synthetic
spectra have appeared recently for the visible range.

Murphy \& Meiksin (\cite{Mur04}), based on Kurucz's ATLAS9 model
atmospheres (Kurucz \cite{Kur93}), have built a high resolution
($\lambda / \Delta\lambda = 250 000$) stellar library over an extended
visible range (3000 to 10000 \AA). The convection zone is treated
using the Mixing Length Theory (MLT) with the overshooting treatment
of Kurucz (cf.\ Castelli et al. \cite{Cas97}). This library provides spectra for 54 values of effective temperature
from 5250 to 50000 K, 11 values of log surface gravity from 0.0 to 5.0
and 19 metallicities from -5.0 to 1.0. They compared their synthetic library with observed spectra (STELIB
library of Le Borgne et al. \cite{LeB03}) for the colours and the Lick
indices and found in general a good agreement.

Still based on Kurucz's models, but with enhanced molecular line lists
and the overshooting option switched off, Munari et al. (\cite{Mun05})
also present a library of synthetic spectra, for a similar wavelength
range (2500 to 10500 \AA). They use a new grid of ATLAS9 model
atmospheres (Castelli \& Kurucz \cite{Cas03}). The effective
temperature of these spectra is contained between 3500 and 47500 K,
the log surface gravity between 0.0 and 5.0 and the metallicity
between -2.5 and 0.5. These spectra are computed at a resolving power
of $\lambda / \Delta\lambda = 500 000$ and then Gaussian convolved to
lower resolution ($\leq$ 20 000). In contrast with Murphy \& Meiksin
(\cite{Mur04}), the predicted energy level lines are not included into
the line lists used to build this library, as Munari et al.
(\cite{Mun05}) favour the spectroscopic use rather than the
photometric use of the synthetic spectra. The addition of these
``predicted lines'' permits to have a better statistical flux
distribution, but individual wavelengths can be wrong by up to 5\%.

We would like to point out here that usually only the lower lying energy
levels of atoms have been determined in the laboratory, particularly for
complex spectra such as those from neutral or singly ionized iron. If only
those transitions were taken into account, the atmospheric line blanketing
computed from such data would be severly incomplete. A lot of weak lines,
possibly unidentified even in the solar spectrum but nevertheless present, would
be missed. This would lead to an overestimation of the ultraviolet flux which
in turn would be ``compensated'' by a lack of flux in the visual (Kurucz
\cite{Kur92}). To avoid this deficiency and to improve the temperature
structure of the model atmospheres, the spectrophotometric flux distribution,
and the photometric colors requires to account for lines for which one or
both energy levels have to be predicted from quantum mechanical calculations.
This has been one of the main goals of the ATLAS9 models of Kurucz
(\cite{Kur92,Kur93}). As the theoretical predictions are accurate to only a few
percent, individual wavelengths can be wrong by up to a few 100~{\AA} in the
visual. Also the line oscillator strengths are sufficiently accurate merely
in a statistical sense. This still permits to improve the total flux within
a wavelength band of a few dozen~{\AA} in the visual, but individual features
do appear in the wrong part of the spectrum. For spectroscopy at higher
resolutions, particularly if the spectrum is rectified or when working within
small wavelength bands, adding the predicted energy level lines ``pollutes''
the theoretical spectrum with extra ``noise''. This has to be avoided in
libraries devoted to automatic fitting procedures or if particular line
features are essential to identify a certain spectral type (lines predicted
for the wrong wavelength will make this more difficult). Hence, with present
atomic data, either choice is only a compromise solution.

By combining three different model atmospheres, the high-resolution stellar library of
Martins et al.\ (\cite{Mar05}) provides the largest coverage in effective temperature
(from 3000 to 50000 K) and log surface gravity (from -0.5 to 5.5). This library, still
in the visible wavelength range (3000 to 7000~\AA), uses the non-LTE model atmosphere
TLUSTY (Hubeny \cite{Hub88}, Hubeny \& Lanz \cite{Hub95}, Lanz \& Hubeny \cite{Lan03})
for $T_\mathrm{eff}\geq 27500 K$, the Kurucz's ATLAS9 model for
$4750\leq T_\mathrm{eff}\leq 27000 K$ and Phoenix/NextGen models
(Allard \& Hauschildt \cite{All95}, Hauschildt et al.\ \cite{Hau99}) which use
spherical symmetry for cooler stars with low surface gravity. A comparison with
observed spectra (from the STELIB library of Le Borgne et al.\ \cite{LeB03} and the
Indo-US library of Valdes et al.\ \cite{Val04}) shows the good agreement of this
theoretical library with observations.

Thus, the visible range is now quite well covered by theoretical libraries, for
photometric use as well as spectroscopic use and for a wide range of physical
parameters. All the comparisons with observations show that these theoretical spectra
can reasonably well mimic real stars, at least at the spectral resolution where the
comparisons were made.

The goal of the present work is to go one step further, exploring the near-infrared
range where few observed fully calibrated and no theoretical libraries are available. This research
takes place in a more general framework which consists in the synthesis of the stellar
population of galaxies hosting active galactic nuclei by an inverse method, described
in Pelat (\cite{Pel97}) and Moultaka \& Pelat (\cite{Mou00}). The H-band provides very
good luminosity discriminators for stars later than K0 (cf Dallier et al.\ \cite{Dal96})
and the particular region 1.57-1.64 $\mu$m of the H-band is clear of strong emission
lines (except Brackett lines). It allows to sample the stellar content of the very
nucleus of Seyfert~1 galaxies, at the contrary of the visible range where the strong
broad emission lines of the active nucleus contaminate so much the spectra of the
galactic inner part that there are too few absorption lines from the stellar component
to synthesize this region.

The lack of stellar observations at medium resolution for the
near-infrared range, especially for super-metallic stars, drove us to
work with theoretical spectra. But the behavior of model atmospheres
and fluxes is not very well known in the infrared.

Decin et al. (\cite{Dec03}) have compared several observed stars with
theoretical spectra computed with the MARCS models (Gustafsson et al.
\cite{Gus75}, Plez et al. \cite{Ple92}) in the range 2.38 to 12 $\mu$m
for the ISO-SWS calibration, at a resolving power $R \simeq 1000$. 
This study points out the difficulties of modelisation due to strong
molecular opacities and the bad accuracy and completeness of the
atomic data in these wavelength ranges.

In this paper, we compute theoretical spectra using the NeMo (Vienna New Model) grid of
atmospheres (Heiter et al. \cite{Hei02}, Nendwich et al. \cite{Nend04}) based on the model
atmosphere code ATLAS9 by Kurucz (\cite{Kur93}, \cite{Kur98}) and
Castelli et al. (\cite{Cas97}), combined with the list of absorption lines
VALD (Vienna Atomic Line Database, Kupka et al. \cite{Kup99}), eventually completed by
molecular data collected by one of us (VT). These models are described in
Sect.~\ref{model}. Synthetic stellar spectra are computed with the code
SynthV (built by VT), as shown in Sect.~\ref{spec}, using the model atmospheres
described in Sect.~\ref{model} as input. Several tests on the input
parameters of the spectra are done in Sect.~\ref{test}.

In a first step of applying our synthesis calculations (Sect.~\ref{visible}), we
compare a set of observed stellar spectra with their corresponding models in the
visible wavelength range (5000 to 9000 \AA) to check the range of validity of the
NeMo grid, exploring the whole range of physical parameters (effective temperature,
surface gravity and metallicity). In a second step, we generate synthetic spectra
in the near-infrared range and compare them with observed ones. The results of
this comparison are described in Sect.~\ref{infrared}, as are tests which
demonstrate that the particular choice of model atmospheres can be expected to
be less important than the set of line lists used for the computation of spectra.
Our conclusions are summarized in Sect.~\ref{conclusions}.

\section{Description of the model atmospheres}\label{model}

NeMo differs from the original grids of model atmospheres based on
ATLAS9 in the treatment of the convective energy transport. It
provides also a higher vertical resolution of the atmospheres and a
finer grid in effective temperature and surface gravity.

This model grid of stellar atmosphere uses convection treatment
without overshooting. The overshooting prescription has been
introduced by Kurucz (\cite{Kur93}, \cite{Kur98}) and modified by
Castelli et al. (\cite{Cas97}). It was supposed to take into account
the change in the temperature gradient of the stable atmosphere layers
near a convective zone due to the overshooting of gas from that zone
into the stellar atmosphere. But this prescription is left aside in
the present work, because, even if the properties of various numerical
simulations are well described and in good agreement when compared to
observations of the Sun, models with overshooting are worse than
models without for other stellar types (see Heiter et al. \cite{Hei02}
for a detailed discussion).

The NeMo grids offer a choice among different convection models. One of them is the
mixing length theory (MLT), with $\alpha = 0.5$. The parameter $\alpha\ $ represents
the ratio between the characteristic length (distance traveled by an element of fluid
before its dissolution) and the scale height of the local pressure. This parameter is
subject to discussion: according to comparisons between observed and computed energy
distributions for the Sun done by Castelli et al. (\cite{Cas97}), $\alpha$ should be
set at least to 1.25, but Van't Veer \& M\'egessier (\cite{vVe96}), using the same
codes and input data as Castelli et al. (\cite{Cas97}), but different observations for
the Sun, found that $\alpha = 0.5$ is required to fit both H$_\alpha$ and H$_\beta$
profiles. Fuhrmann et al. (\cite{Fur93}) were the first to notice that a value of 0.5
for the parameter $\alpha$ is needed to reproduce the Balmer line profiles of cool
dwarf stars. In addition, this parameter has to span a large domain (from 1 to 3) to
reproduce the red giants (Stothers \& Chin \cite{Sto97}). The alternative convective
models available in the NeMo grids are of "Full Spectrum Turbulence" (FST) type.
Introduced by Canuto \& Mazzitelli (\cite{Can91}, \cite{Can92}; thereafter
CM model) and Canuto, Goldman \& Mazzitelli (\cite{Can96}; thereafter CGM model),
these models avoid the one-eddy approximation of MLT. In addition, both models were
suggested to be used with a scale length different from the usual multiple $\alpha$
of the local pressure scale height (see Heiter et al. \cite{Hei02} for further details).

The latter models were introduced in NeMo to allow a choice among different
treatments of the internal structure of the stars, depending on the aim of the
model computation and its underlying assumption of how to describe the
convective energy transport within the limitations of a simple convection model
(using only algebraic rather than differential equations). 

Two levels of vertical resolution are also offered and hence we can
either work with 72 or 288 layers. The MLT models are computed with 72
layers, CM models with 288 and CGM ones are computed for both values.

The metallicity of the model atmopheres covers a large range between
-2.0 and +1.0 dex and have 13 different values. This range of
metallicity is enough for our purpose. The super metal rich stars, in
particular, are represented with five different levels of metallicity
(+0.1, +0.2, +0.3, +0.5 and +1.0 dex) reaching the highest possible
value for a real star.

NeMo provides model atmospheres for effective temperatures between
4000K and 10000K, by successive steps of 200K; for lowest
temperatures, the model atmospheres computed with ATLAS9 become
inadequate, mainly because of the molecular opacities which become
very important for cool stars. The MARCS6 models (Gustafsson et al.
\cite{Gus75}, Plez et al. \cite{Ple92}), more dedicated to the cool
stars, handle this problem with a more complete treatment of molecular
opacity.

The available values for the surface gravity (log g) of the stellar
atmospheres in the NeMo grid span a range from 2.0 to 5.0 with steps
of 0.2. It is bounded at 2.0 owing to the plane-parallel approximation
used in ATLAS9; for lower values of log g, spherically symmetric
geometry should be used
instead (cf.\ Hauschildt et al.\ \cite{Hau99} and Baraffe et al.\ \cite{Bar02}).

Other models, working with the appropriate geometry like MARCS6 or
Phoenix/NextGen (Allard \& Hauschildt \cite{All95}, Hauschildt et al.
\cite{Hau99}) are necessary for these small values. Indeed, MARCS6,
whose main purpose is to model cool stars, uses also the approximation
of spherically symmetric geometry to reproduce supergiants and the
cool giants stars, which have a low surface gravity (Plez et al.\
\cite{Ple92}).  NextGen models, like ATLAS9, assume LTE and
plane-parallel geometry for dwarf stars, but a spherical symmetry is
used for low-gravity giant and pre-main sequence stars (log g $<$ 3.5,
see Hauschildt et al.\ \cite{Hau99}).

Contrary to NeMo and MARCS6, NextGen can use a non-LTE model for high
temperature stars. But using NLTE does not improve significantly the
modelisation of our observed stars, as NLTE effects begin to occur
only from 7000 K to higher effective temperature (Hauschildt et al.
\cite{Hau99}), but are still small to at least 10000 K. Moreover,
NextGen does not reproduce well enough the individual lines owing to
the treatment of atomic and molecular lines with a direct opacity
sampling method. Indeed, working with opacity distribution functions,
like in ATLAS9, would ask too much computer resources when using NLTE
calculations (Hauschildt et al. \cite{Hau99}). In addition to that,
too few layers are used in the published models to describe the bottom
part of the photosphere.

However, NextGen could be an alternative to ATLAS9 type model atmospheres in
a next step of our project, for generating spectra of stars with a log g below
2.0 (for which spherical symmetry is needed) and/or an effective temperature
above 10000 K.

Bertone et al. (\cite{Ber04}) have compared both ATLAS9 and NextGen models to
observations in the visible range along the whole spectral-type sequence. The
conclusions of this work are that both models reproduce very well the spectral energy
distribution of F type stars and earlier but this good agreement decreases at lower
temperature, especially for K stars, owing to the lack of molecular treatment in those
models. ATLAS9 provides a better fit, in general, from B to K type stars but as said
previously NextGen is more suitable for M stars, due to the use of spherical geommetry
for the giants and a more complete molecular line opacity. However, Martins et
al.\ (\cite{Mar05}) note that this comparison is made with a previous generation
of NextGen models, using for example a mixing length parameter of 1 instead of
2, preferred by hydrodynamic models. This is also true for the ATLAS9 models, as
Bertone et al.\ (\cite{Ber04}) did not use the latest versions of ATLAS9, including
new opacity distribution functions (Castelli \& Kurucz \cite{Cas03}), computed with
more up-to-date solar abundances and molecular contributions than the previous one.

Thus, a new comparison with observations in the visible range is not
unnecessary. Moreover, an extensive comparison of spectra based on the
NeMo grid of model atmospheres for the entire range of A to early M stars
including both dwarfs and giants has not been done before. We hence begin
our comparison with observations in the visual before proceeding to the
infrared. The implications of changing abundances or the description of
convection at spectral resolutions relevant for studies of galaxies
are included as part of the discussion of our comparisons.

\section{Obtaining a theoretical spectrum}\label{spec}

First of all, we downloaded the model atmospheres corresponding to the stellar types
wanted from the NeMo website (http://ams.astro.univie.ac.at/nemo/). The models are
classified according to the convection model (CM, CGM or MLT) and to the number of
layers representing the atmosphere. The model CGM with 72 layers is detailed enough
for our purpose. Models with 288 layers are used only for specific applications like
the calculation of the convective scale length in stellar interior models
(Heiter et al.\ \cite{Hei02}). Reduced to the medium resolution of our observations,
both computations of a model with 72 and 288 layers respectively give similar spectra.

The next parameter to determine is the microturbulence velocity. For cool dwarf
stars, this velocity is low: about 0--1~km/s, but the value is increasing
towards higher luminosities, reaching values as high as 5~km/s (Gray \cite{Gra92}).
A few stars do not follow this rule: hot stars, like B and O type, have
a negligible microturbulence velocity and some specific types of A stars can
either have a null velocity (Ap type stars, for them magnetic field effects are
important instead) or a velocity of 4~km/s (Am type stars). As the
microturbulence velocity has only a small influence on the overall shape of the
spectra and on the line profiles at our spectral resolution, we can use a common
value of 2~km/s for comparison with all our stellar spectra, composed by A to
early-M type dwarf and F to K type giant stars, as 2~km/s is a good compromise
for these stars (Gray \cite{Gra92}).

Then, the three main physical parameters of the star have to be
chosen. The metallicity, the effective temperature and the surface
gravity of the theoretical stellar spectrum should correspond as good
as possible to the observed star to be compared. Therefore, once the
stellar characteristics are determined, the nearest set of parameters
(T, log g, Z) in the NeMo grid is taken.

The metallicity is taken from Nordstroem et al. (\cite{Nor04}), Cayrel
de Strobel et al. (\cite{Cay01}) and Barbuy \& Grenon (\cite{Bar90})
when available for individual stars of the sample, otherwise assumed
to be solar.

Nordstroem et al. (\cite{Nor04}) have determined the effective
temperature of most of the dwarf stars in our sample, for other stars
the effective temperature is assumed, as well as the surface gravity,
according to their spectral type from the corresponding values of
temperature and gravity as given in Schmidt-Kaler (\cite{Sch82}) and
Gray (\cite{Gra92}).

Once the most suitable model atmosphere is determined, we can generate
a theoretical flux calibrated spectrum with the code SynthV (by VT).
This code requires several input parameters such as the wavelength
range for which the spectrum will be computed, as well as the
wavelength step. This step has to be small enough compared to 2.5~\AA\
as the desired opacity in each wavelength point includes absorption
from all nearest lines within 2.5 \AA, so large wavelength steps would
give wrong results. We take 0.1~\AA\ in the visible and in the
infrared range. Then, we can enter a rotation profile for the star, if
needed, and indicate a list of absorption lines to be used.  For our
work, we take the Vienna Atomic Linelist Database (VALD), completed by
several molecular line lists (C$_2$, CN, CO, H$_2$, CH, NH, OH, MgH,
SiH, SiO, TiO, H$_2$O \footnote{file VColl\_molec.lns built by VT
    from Kurucz' CDROMs 15, 24, 25, and 26, see Kurucz \cite{Kur93b}
    and \cite{Kur99}; the file is available from V. Tsymbal upon
    request}).  The line profiles are approximated by a Voigt
function. SynthV also provides the possibility to change individual
abundances.

The final theoretical spectrum has to be reduced to the same
resolution and the same sampling as the observed spectrum for further
comparisons. So, the calculated spectrum is Gaussian smoothed and
resampled by Fourier interpolation to the same step as the observed
spectrum.

\section{Testing parameters}\label{test}

The determination of the physical parameters of the observed stars is
not as acurate as we would like.  So it is necessary to investigate
the nearest values of T$_\mathrm{eff}$ / log g / Z of the grid. The
chemical abundances can also be changed; as these abundances are not
very well determined, it is crucial to notice how a variation of the
abundance of one element modifies the spectrum.

The most important change is caused by the temperature. Indeed, the step of
{200 K} as was chosen for the grid computation is still quite large for our 
purpose and a deviation of this range can be dramatic for the slope of the
spectrum. The coldest stars (M, K and even G type) are the ones most affected
by a change of 200~K.

\begin{figure*}
\centering
  \includegraphics[angle=-90,width=12cm]{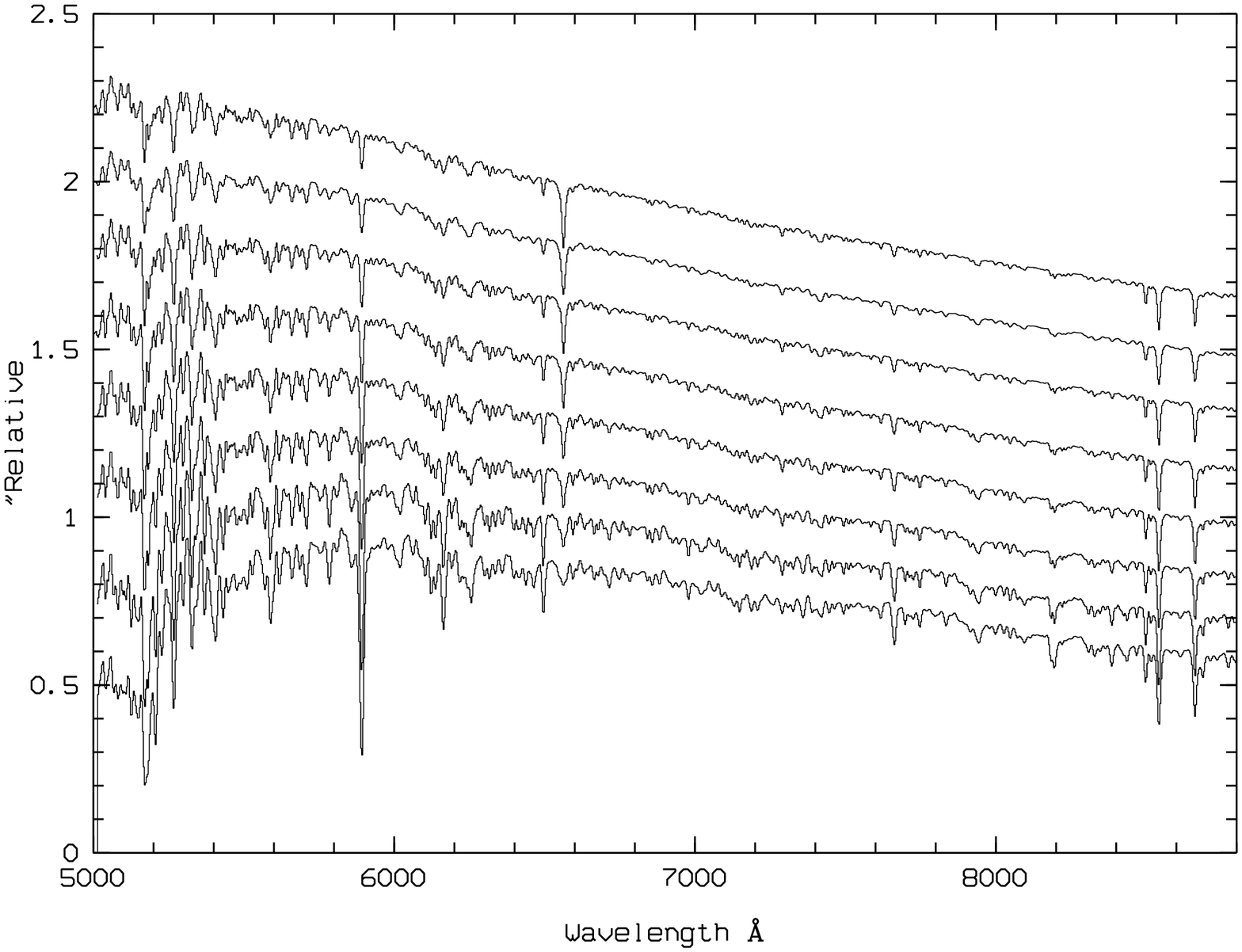}
  \caption{Results of a variation of temperature in the visible range
    for dwarf stars. From the top to the bottom: T=6000K to 4600K with
    a step of 200K between two spectra, arbitrarily shifted by a
    constant value for the purpose of clarity.}
  \label{vis-dw}
\end{figure*}

\begin{figure*}
\centering
  \includegraphics[angle=-90,width=12cm]{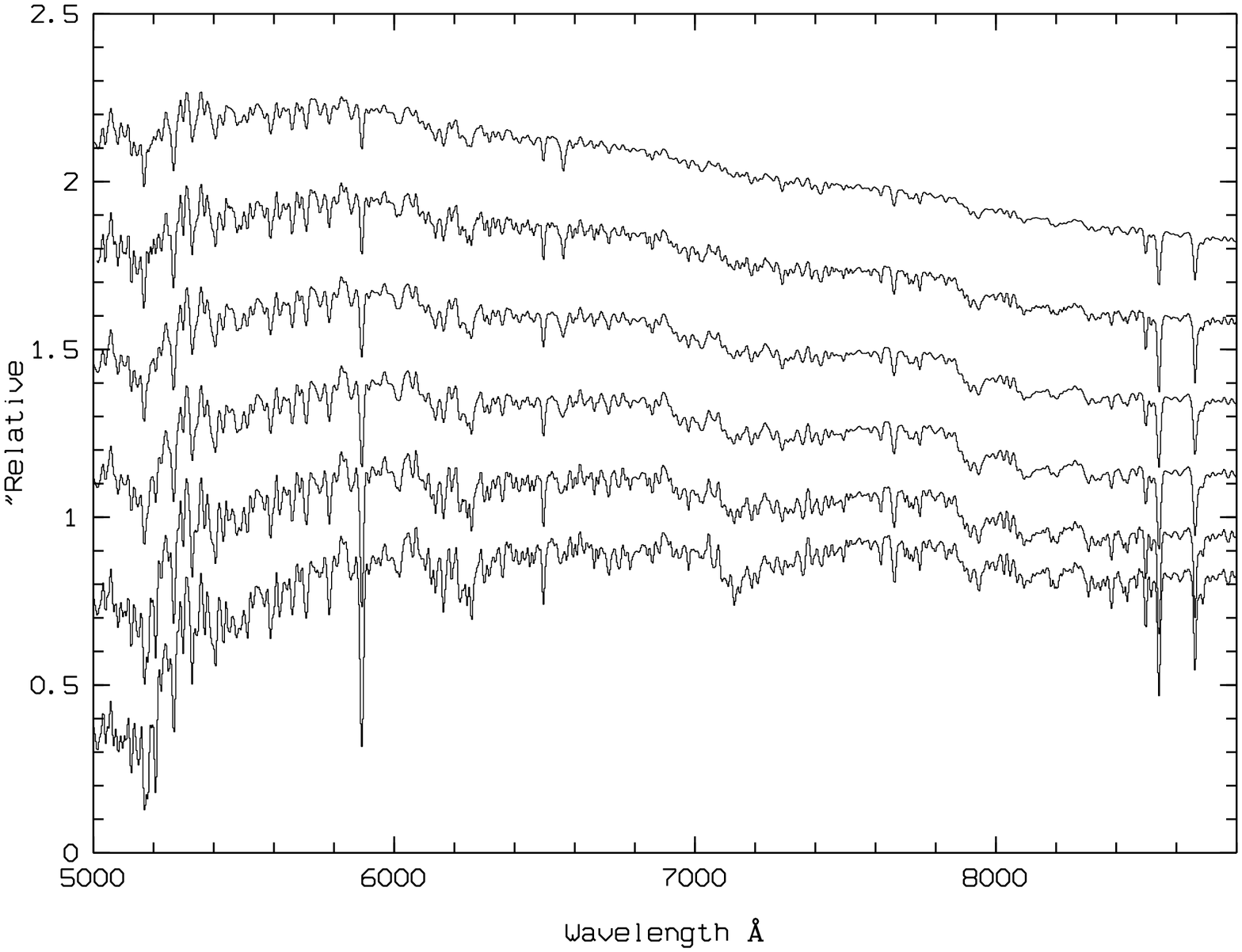}
  \caption{Results of a variation of temperature in the visible range
    for giant stars. From the top to the bottom: T=5000K to 4000K with
    a step of 200K between two spectra, arbitrarily shifted by a
    constant value for the purpose of clarity.}
  \label{vis-gi}
\end{figure*}

\begin{figure*}
  \centering
  \includegraphics[angle=-90,width=12cm]{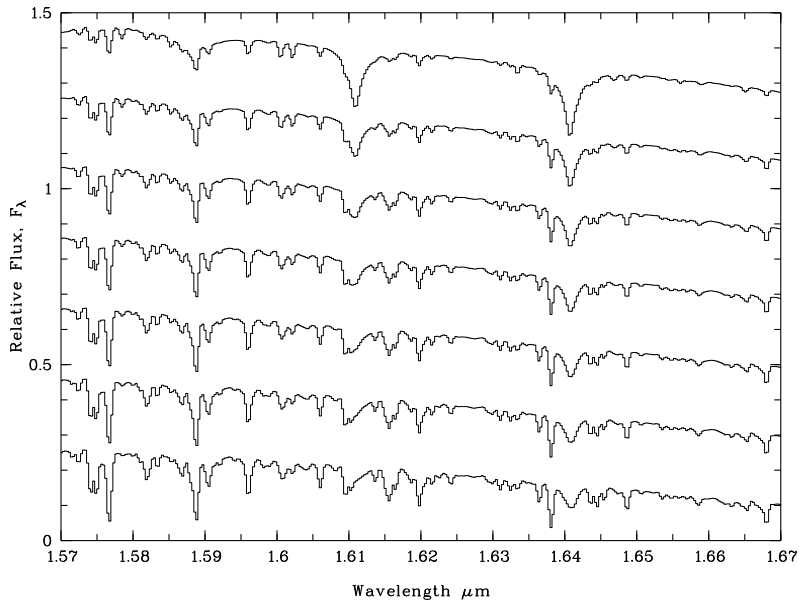}

  \caption{Results of a variation of temperature in the infrared range
    for dwarf stars. From the top to the bottom: T=7200K, T=6400K and
    T=6000K to 5200K with a step of 200K, arbitrarily shifted by a
    constant value for the purpose of clarity.}
  \label{ir-kgi}
\end{figure*}

Fig. \ref{vis-dw} and \ref{vis-gi} show the evolution of the spectra
with temperature in the visible range for the dwarfs and the giants,
respectively, and Fig. \ref{ir-kgi} shows this evolution for dwarf
stars in the infrared range. We can see that for intermediate
temperature, there is mainly a difference in continuum. But for the
extreme values, the modification of the spectrum is more dramatic, as
it affects also the absorption line features.

A change in log g causes a variation of the line profiles. This
parameter is not very well known for observed stars. So it is
important to test different values around a first guess. Even if the
variation caused by a gap of 0.2 in log g is not very important, it
can improve the comparison.

The metallicity has an influence on the slope of the continuum:
increasing the metallicity of a theoretical spectrum has a similar
influence on the continuum as decreasing the temperature (see e.g.
Ram\'irez \& Mel\'endez, 2005, for a detailled discussion). A
variation of metallicity has also a clear influence on the strength of
the absorption lines.

The variation of individual abundances can also cause some changes in the
spectra. SynthV uses by default solar abundances of Anders \& Grevesse
(\cite{And89}). But some studies more recent like Kurucz (\cite{Kur93}) or Holweger (\cite{Hol01}) give different
values for various elements (like He, Fe, O, C, N...). These changes can be
quite important (up to 0.2 dex for Fe). A simple test to probe the influence of
different abundances was made comparing two spectra assuming the same physical
parameters, except for the solar abundance of Holweger (\cite{Hol01}) in one
case and the one used by Kurucz (\cite{Kur93}) in the other. The result shows only slight differences,
and the comparison with our observations cannot determine whether one is better
than the other. For our work, we chose the values given by Holweger (\cite{Hol01}).

In order to fit metallic stars, it is also important to check the
influence of the modification of an individual element abundance on
the synthetic spectrum. A change of the individual abundances of O and
C leads to a modification of OH and CO line strengths, respectively.
Indeed, when the ratio C/O increases, more CO molecules will be
formed; on the other hand, when it decreases, more oxygen will be left
to form OH molecules (see Decin et al. \cite{Dec00}).

For the hottest stars, it is important to take into account rotational
velocity.  At the medium resolution of our observations, a convolution
with a Gaussian profile is good enough to reproduce this effect.
Hence, we do not need to include a more accurate description of the
change in the profile of the lines caused by the rotational velocity.
Consequently, even though SynthV allows to compute spectra with a
rotational velocity profile for the lines, we compute the spectra
without rotational velocity in order to save computation time and
convolve them afterwards with a gaussian profile.

Additional tests have been made with a different microturbulence
velocity for a very cool star (4 km/s for a M0V-type spectrum in the
visible range) and a different number of layers for the convection
model (288 instead of 72 for the same CGM model and the same physical
parameters). At the resolution of the observed samples, these
modifications do not lead to any difference.

\section{Results in the visible range}\label{visible}

Although our goal is to explore spectra in the infrared wavelength range, a study of
the behavior of the NeMo model atmospheres and spectra in the visible range provides
us a good indication of their reliability as a function of the physical parameters of
the stars.

\subsection{Observations in the visible}

18 spectra of observed stars (A to M dwarfs and G to K giants)
corresponding to the range of the parameters in the NeMo grid have
been compared to theoretical stellar spectra. This sample of
observed spectra at a resolving power of R $\simeq$ 600 is taken from
the stellar library used by Boisson et al. (\cite{Boi00}). One part
of these observations comes from the stellar library of Silva \&
Cornell (\cite{Sil92}), made at the KPNO with the MARK III spectrograph,
they cover the wavelength range 3500-9000\AA. These spectra
are, for most of them, a mean-value of several stars of nearby
spectral type. The name of these stars and the mean associated
spectral types as given by Silva \& Cornell are listed in Table
\ref{list_vis}.  The remaining of the library, mainly supermetallic
stars, were observed by Serote Roos et al. (\cite{Ser96}) at the CFHT
with the Herzberg spectrograph and at OHP with the Aurelie
spectrograph. The spectral range is limited to 5000-9000\AA.
The name, spectral type (or associated mean-spectral type) and
parameters of these stars are listed in Table \ref{list_vis}.

\begin{table*}
\caption{List of observed stars in the visible, with the values of the parameters
taken from the mean-values listed by Gray (\cite{Gra92}) and Schmidt-Kaler
(\cite{Sch82}) or from (1) Nordstroem et al.\ (\cite{Nor04}),
(2) Cayrel de Strobel et al.\ (\cite{Cay01}) and (3) Barbuy \& Grenon (\cite{Bar90}).
In the last column, when no information on metallicity is available, a tick mark
replaces it. The quantity $<v\sin i>$ is given in km/s.}
\label{list_vis}
\centering
\begin{tabular}{|p{4.0cm} c c p{2.8cm} c p{2.5cm}|}
\hline
Name & Spectral Type & $<v\sin i>$ & $T_\mathrm{eff}$(K) & log g & [Fe/H] \\
\hline
HD116608,HD190785,HD124320 & A1-3 V & $145$ & $8900$ & $4.2$ & - \\
\& HD221741& & & & & \\
HD88815 & F2 V & $90$ & $7244^1$ & $4.3$ & $-0.13^1$ \\
HD187691, HD149890 & F8-9 V & $7$ & $6026^1$, $5902^1$ & $4.4$ & $0.07^2$, $-0.44^1$ \\
HD121370 & rG0 IV & $5$ & $5957^1$ & $4.4$ & $+0.27^2$ \\
HD38858 & G4 V & $3$ & $5636^1$ & $4.5$ & $-0.26^1$ \\
HD161797 & rG5 IV & $3$ & $5700$ & $4.5$ & $+0.23^2$ \\
HD149661,HD151541,HD33278, & G9K0 V & $2$ & $5176^1$, $5236^1$,
   $5300$, & $4.5$ & $0.01^1$, $-0.36^1$, - \\
HD23524,SAO66004,SAO84725 & & & $5200^1$, $5300$, $5300$& &  $-0.49^1$, -, -\\
HD93800 & rK0 V & $2$ & $5250$ & $4.5$ & $+0.43^3$ \\
HD39715 & rK3 V & $1$ & $4850$ & $4.6$ & $+0.33^3$ \\
HD36395 & rM1 V & $1$ & $3850$ & $4.6$ & $+0.6^2$ \\
\hline
HD15866, HD25894, HD2506 & G0-4 III & $10$ & $5500$ & $3.0$ & - \\
HD163993 & wG8 III & $3$ & $4950$ & $2.7$ & $-0.1^2$ \\
HD72324 & G9 III & $3$ & $4900$ & $2.7$ & - \\
HD33506, HD112989 & rG9K2 III & $2$ & $4700$ & $2.6$ & $+0.14^2$ \\
SAO76803 & K2 III & $2$ & $4500$ & $2.5$ & - \\
HD181984 & rK2 III & $2$ & $4500$ & $2.5$ & $+0.1^2$\\
HD176670 & rK3 III & $1$ & $4300$ & $2.2$ & $-0.03^2$ \\
HD154733, HD21110 & K4 III & $1$ & $4000$ & $2.0$ & $-0.14^2$,- \\
\hline
\end{tabular}
\end{table*}

The atmospheric bands are removed from the observed spectra to compare with the theoretical spectra.

\subsection{Comparisons}

\begin{figure*}
\centering
\includegraphics[angle=-90,width=12cm]{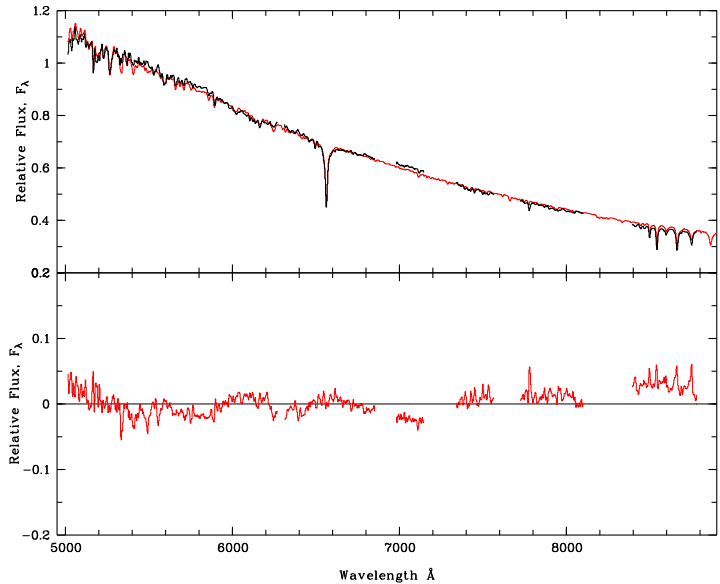}
\caption{The observed spectrum (in black) is from a F2V type star (HD 88815), the
         theoretical one (in grey) is computed with the following parameters:
		 T=7200K, log g=4.0, [M/H]=-0.1. Atmospheric bands are removed from
		 the observed spectrum. The residual between the two spectra is 
		 (theoretical~flux~-~observed~flux)/theoretical~flux.}
\label{v-f2v}
\end{figure*}

When the physical parameters (see Table \ref{list_vis})
are known, we took the model with the nearest values, otherwise we used the mean values
according to the spectral type as starting points and investigated the nearby values to
find the best agreement between the observed and the computed spectra. The theoretical
spectra are computed with a wavelength step of 0.1~\AA, corresponding to a resolution
of 60000 at 6000~\AA, then Gaussian smoothed to the resolution of the observed spectra.
By Fourier interpolation, we reduce the computed spectra to the same wavelength step
as the observations. Spectra are normalized to 1 in the range 5440-5460\AA. When the star
has a rotational velocity, we convolve the corresponding computed spectrum with
a Gaussian of the same velocity.

\begin{figure*}
\centering
\includegraphics[angle=-90,width=12cm]{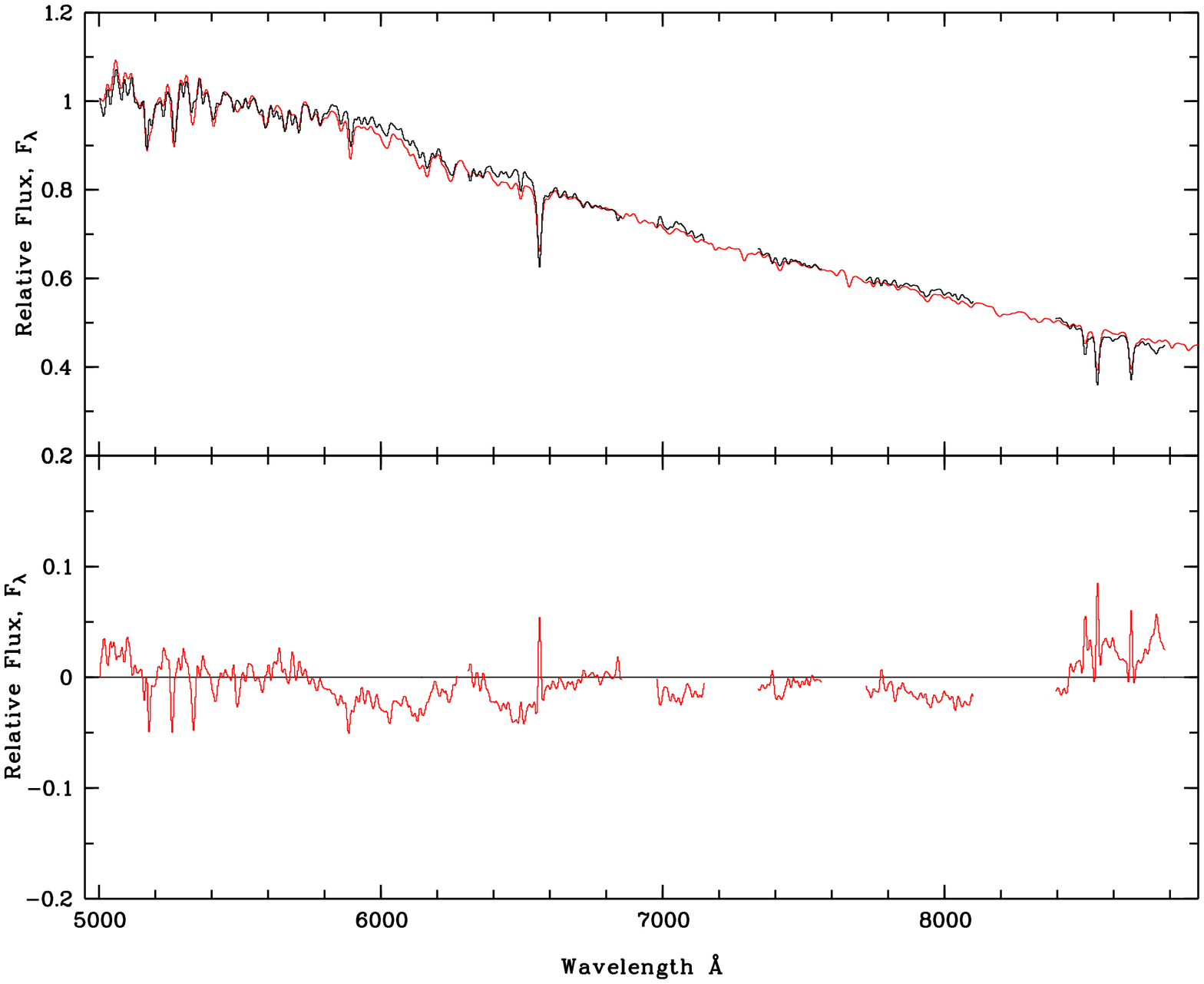}
\caption{The observed spectrum (in black) is from a G0IV type star (HD
  121370), the theoretical one (in grey) is computed with the
  following parameters: T=6000K, log g=4.4, [M/H]=+0.3.}
\label{v-g0iv}
\end{figure*}

\begin{figure*}
\centering
\includegraphics[angle=-90,width=12cm]{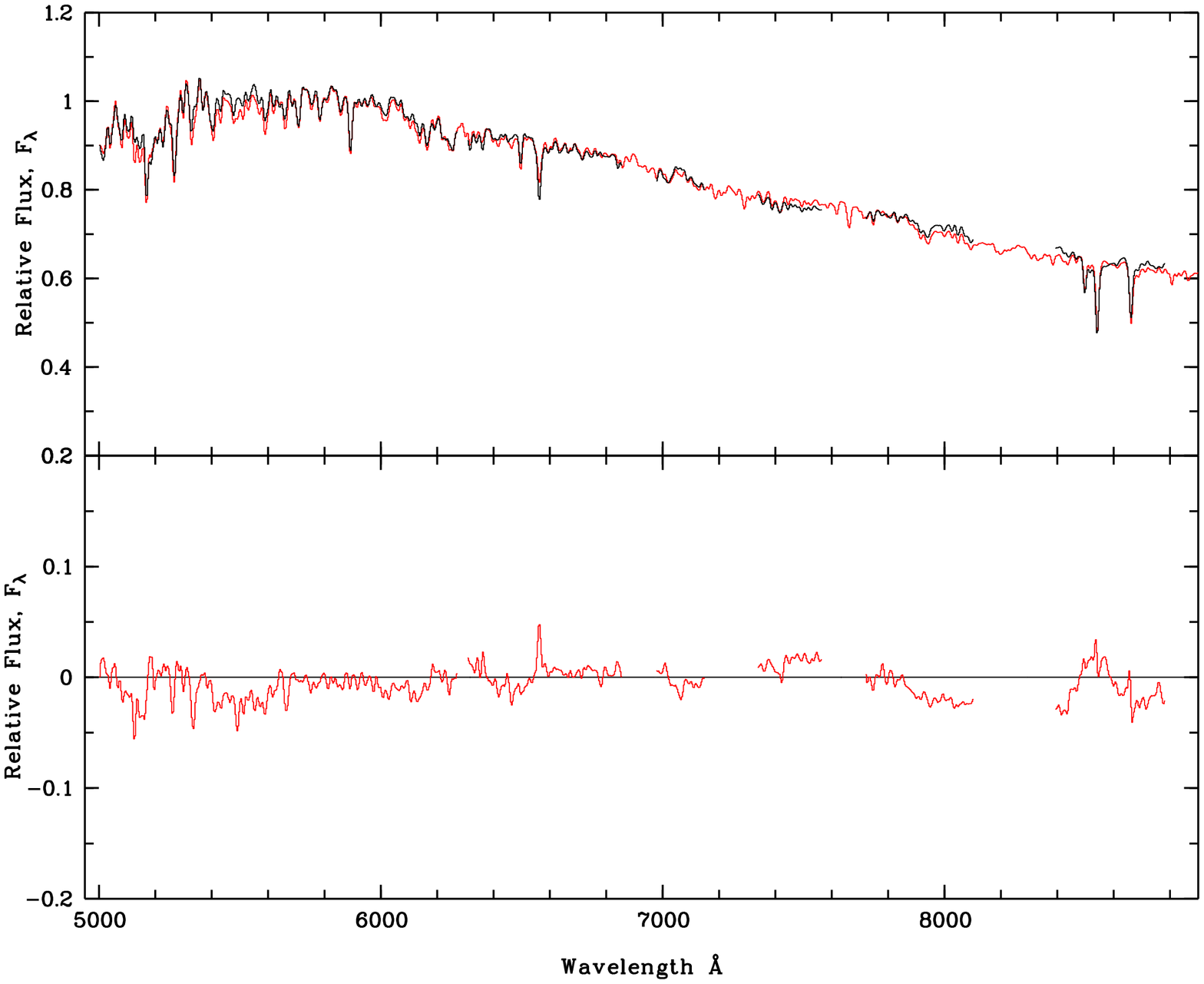}
\caption{The observed spectrum (in black) is from a G8III type star
  (HD 163993), the theoretical one (in grey) is computed with the
  following parameters: T=5000K, log g=2.8, [M/H]=-0.1.}
\label{v-g8iii}
\end{figure*}

\begin{figure*}
\centering
\includegraphics[angle=-90,width=12cm]{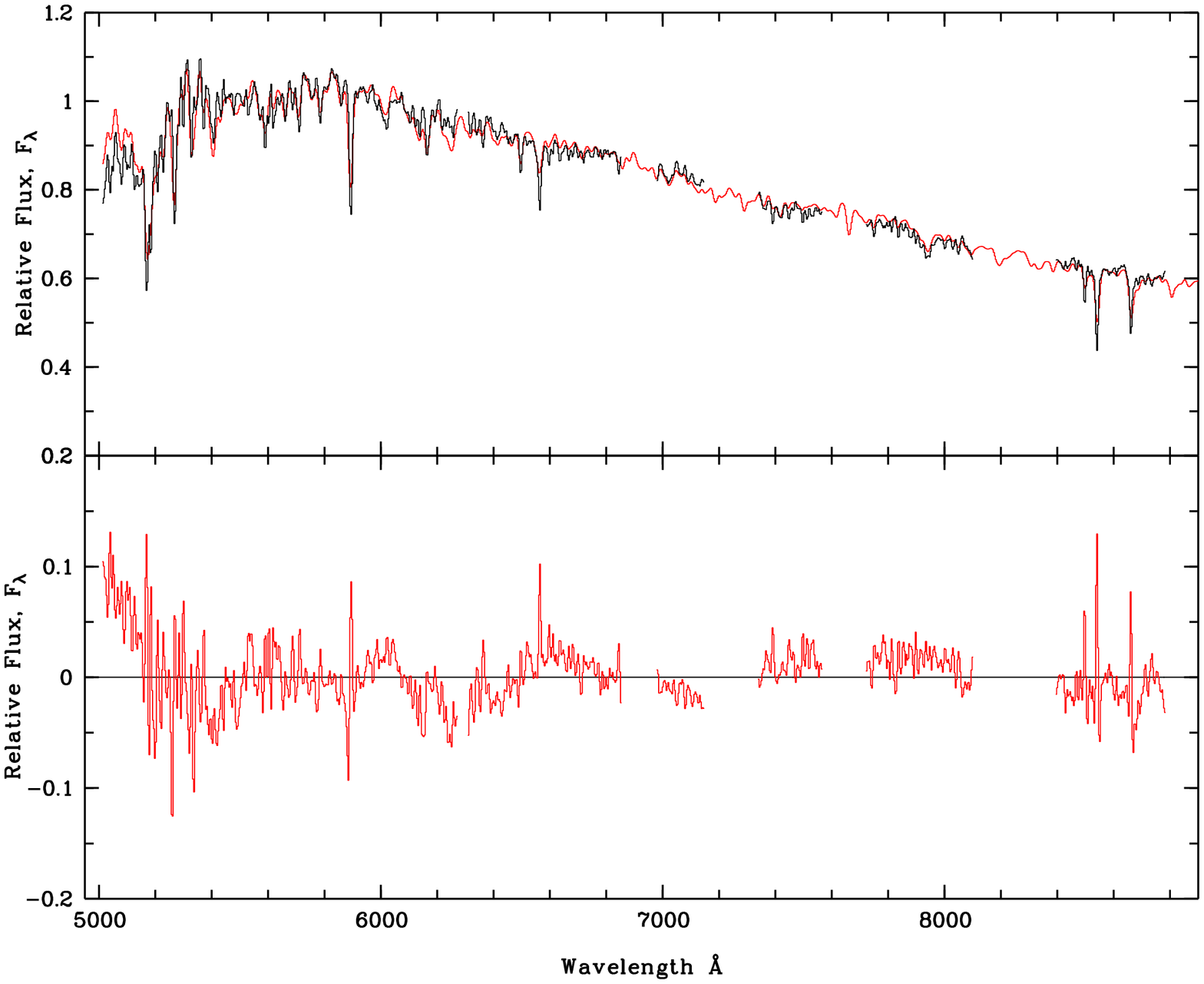}
\caption{The observed spectrum (in black) is from a K0V type star (HD
  93800), the theoretical one (in grey) is computed with the following
  parameters: T=5200K, log g=4.4, [M/H]=+0.5.}
\label{v-k0v}
\end{figure*}

The agreement between the observed and computed spectra is
satisfactory for effective temperatures ranging 4600 to 9000 K, 9000 K
corresponding to the highest temperature for the stars composing our
sample. For these spectra, the main discrepancies, which consist in
differences in the slope of the blue extremity of the continuum, can
be explained by the difficulty to have a good flux calibration at the
wavelength ends of the observational data (in particular at 5000\AA\
where strong MgI, MgH and FeI absorptions are present). Various
examples of comparisons for these stars are shown in
Fig.~\ref{v-f2v}-\ref{v-k0v}.  The spectral type of the observed stars
and the physical parameters used to compute the theoretical spectra
are noticed for each figure.

\begin{figure*}
\centering
\includegraphics[angle=-90,width=12cm]{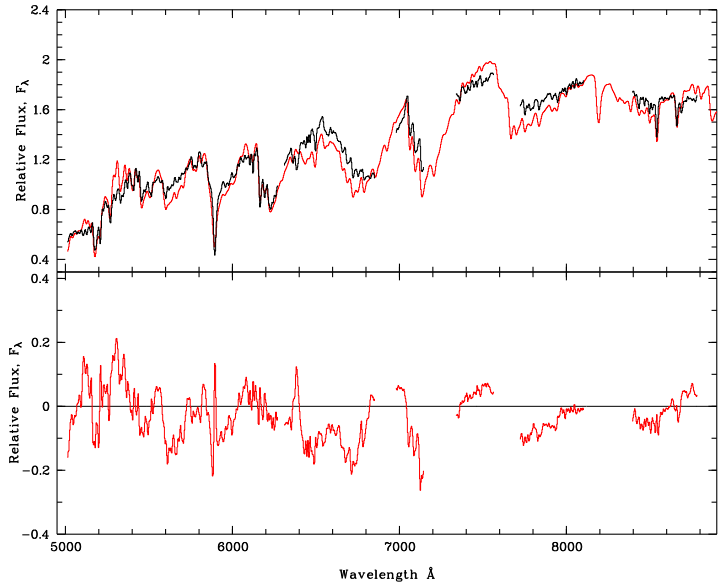}
\caption{The observed spectrum (in black) is from a M1V type star (HD
  36395), the theoretical one (in grey) is computed with the following
  parameters : T=4000K, log g=4.6, [M/H]=+1.0. The scale for the
  residual is twice the scale of the previous figures.}
\label{v-m1v}
\end{figure*}

\begin{figure*}
\centering
\includegraphics[angle=-90,width=12cm]{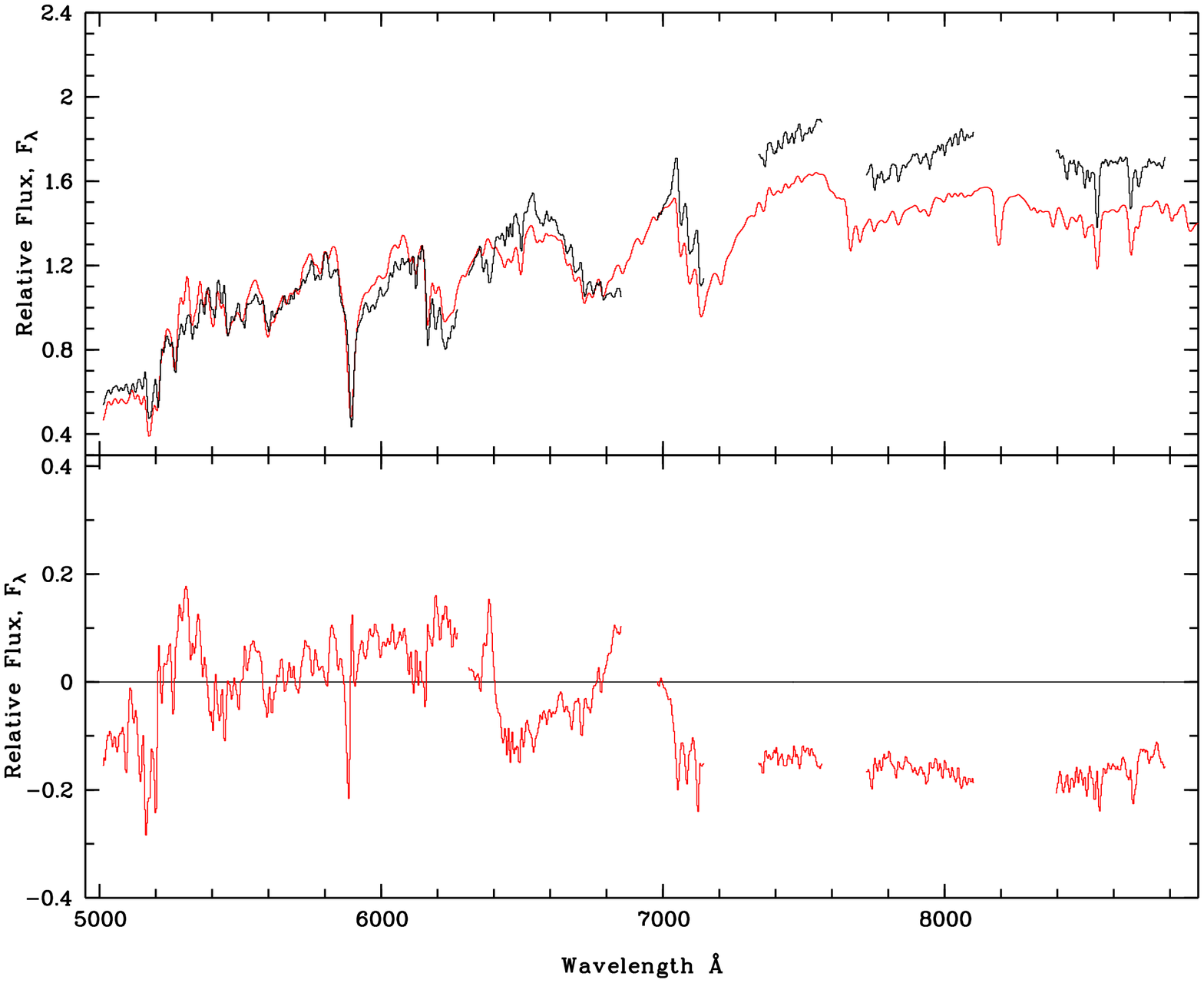}
\caption{Same as Fig. \ref{v-m1v}, with [M/H]=+0.5 for the theoretical spectrum.}
\label{v-m1v_5}
\end{figure*}

Three stars have a temperature below 4400 K, one dwarf and two giants. The M dwarf
star (HD 36395) is particular as it has a very high metallicity. It is the most
metallic star of Cayrel de Strobel's catalog.\footnote{[Fe/H]=+0.6 dex} Indeed, to
fit correctly the observations, we need to set the metallicity of the theoretical star
as high as possible (Fig.~\ref{v-m1v}), but a metallicity of [M/H]=+1.0 dex is not
realistic. Fig.~\ref{v-m1v_5} shows that [M/H]=+0.5 dex is not sufficient. However,
recently Woolf \& Wallerstein (\cite{Woo05}) have found a temperature of 3760~K (instead of 3850~K) and
a metallicity of [Fe/H]=+0.2 dex for this star. So the discrepancies may be simply
due to the difference in temperature, out of reach for NeMo.

\begin{figure*}
\centering
\includegraphics[angle=-90,width=12cm]{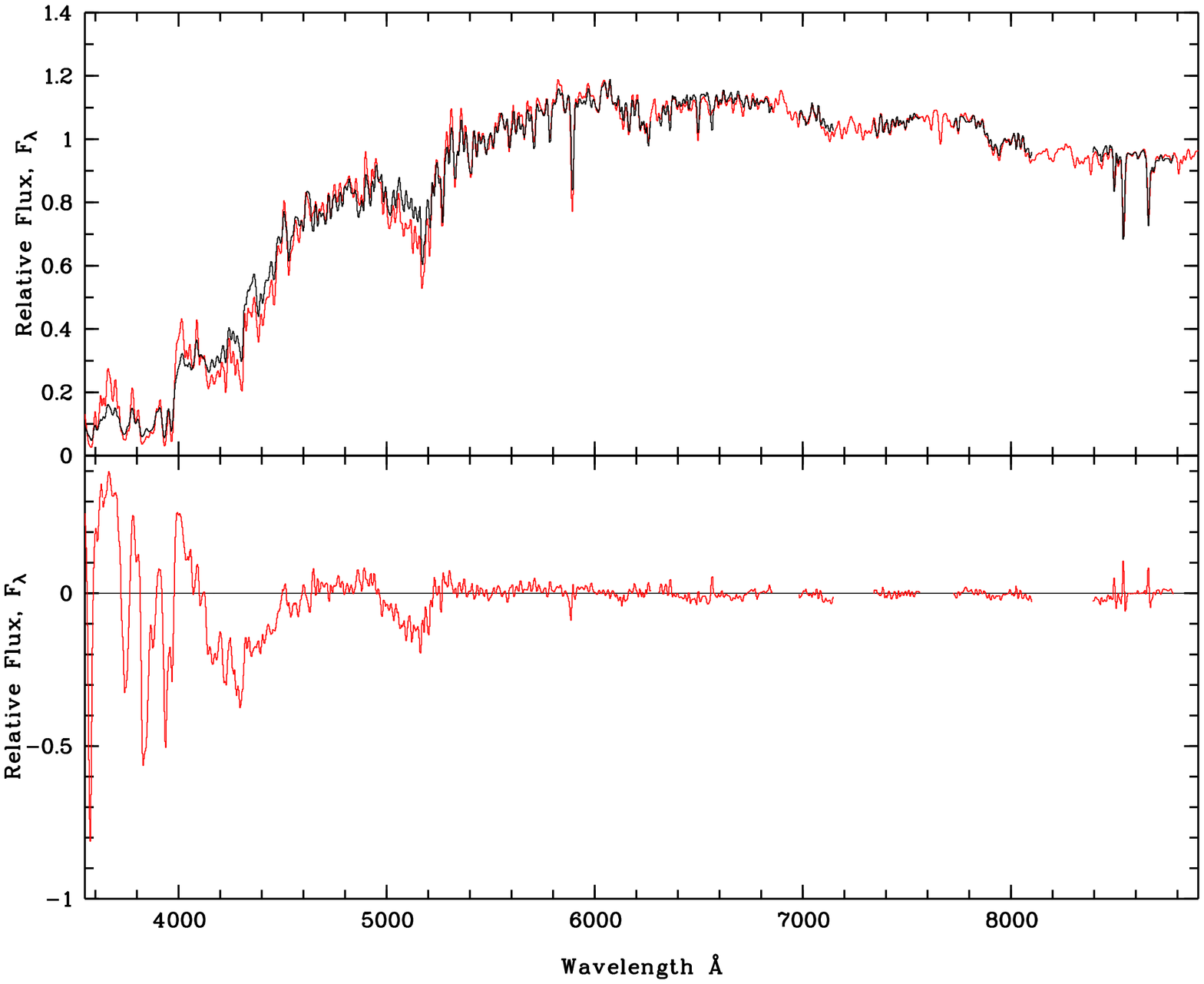}
\caption{The observed spectrum (in black) is the mean-value of two K4III type stars
         (HD 154733 and HD 21110), the theoretical one (in grey) is computed with the
		 following parameters: T=4000K, log g=2.0, solar metallicity. The scale of
		 this plot is not the same as for the previous figures.}
\label{v-k4iii}
\end{figure*}

The two others are K giant stars. First it is worth to say that at
this low temperature, a difference of 200 K causes a variation of the
slope of the continuum more important than for the highest
temperatures. Thus, the continuum of the K3III observed star can
neither be correctly fitted by a computed spectrum with a temperature
of 4400 K nor with 4200 K. The first one is too blue and the second
one is too red.  Another difficulty comes from the Na line at 5894
\AA\ and the MgH band computed too strong compared to the
observations. These absorption features are very sensitive to a
variation of temperature and a too strong value means that the
temperature is too low, but increasing the temperature would lead to a
continuum too blue. This is particularly clear for the K4III star,
which lies on the boundaries of the NeMo grid for temperature and
gravity (T=4000 K, log g=2.0), as we can see on Fig. \ref{v-k4iii}.
This star comes from the stellar library of Silva \& Cornell (1992)
whose original spectra extend down to 3500\AA.  The agreement for the
slope of the continuum is satisfactory, but the computed NaI line,
  the MgH band as well as other lines and molecular bands such as e.g.
  CaII, CN, Gband are by far too strong.  If we set the temperature at
  4200~K, which is still reasonable for this kind of star, the
  accordance for the lines and molecular bands would be better, but
  the slope of the observed continuum would be too red.

The hypothesis of plane-parallel geometry of the model begins to
become unrealistic (low log gravity) and the molecular opacities,
which are not taken sufficiently into account in ATLAS9, become
important for these cool stars.

\section{Results in the infrared range}\label{infrared}

We can immediatly say that there are, by far, more discrepancies
between the computed and the observed spectra in the infrared than in
the visible range, even at the medium resolution we have.

\subsection{Observations in the infrared}

The observed spectra for this wavelength range come from Meyer et al. (\cite{Mey98})
and Boisson et al. (\cite{Boi02}). The Meyer's ones are observations at a resolving
power of R $\simeq$ 3000 at 1.6 $\mu$m with the KPNO Mayall 4 m Fourier Transform
Spectrometer; these spectra have to be calibrated in flux. The stars from Boisson
et al. (\cite{Boi02}) come from the ISAAC spectrograph, mounted on the VLT telescope,
at a resolving power of R $\simeq$ 3300 at 1.6 $\mu$m. From these samples, we selected
23 stars matching the available parameters of NeMo (A to M dwarfs and F to K giants),
listed in Table \ref{list_ir}.

\begin{table*}
\caption{List of observed stars in the infrared, with the values of the parameters
         taken from the mean-values listed by Gray (\cite{Gra92}) and Schmidt-Kaler
		 (\cite{Sch82}) or from (1) Nordstroem et al. (\cite{Nor04}),
		 (2) Cayrel de Strobel et al. (\cite{Cay01}) and (3) Barbuy \& Grenon
		 (\cite{Bar90}). In the last column, when no information on metallicity
		 is available, a tick mark replaces it.}
\label{list_ir}
\begin{tabular}{|l c c c c c|}
\hline
Name & Spectral Type & $<v\sin i>$(km/s) & $T_\mathrm{eff}$(K) & log g & [Fe/H] \\
\hline
HD159217 & A0 V & $150$ & $9700$ & $4.3$ & $-$ \\
HD27397 & F0 IV & $120$ & $7100$ & $4.3$ & $-$ \\
HD48501 & F2 V & $9$0 & $6850$ & $4.3$ & $+0.01$ \\
HD26015 & F3 V & 65 & $6776^1$ & $4.3$ & $+0.11^1$\\
HD30606 & F6 V & 10 & $6152^1$ & $4.4$ & $-0.01^2$ \\
HD98231 & F8.5 V & 7 & $5794^1$ & $4.4$ & $-0.35^2$ \\
HD112164 & rG1 V & 5 & $5768^1$ & $4.4$ & $+0.24^2$ \\
HD10307 & G1.5 V & 4 & $5781^1$ & $4.4$ & $-0.04^2$ \\
HD98230 & G2 V & 4 & $5794^1$ & $4.5$ & $-0.34^2$ \\
HD106116 & rG4 V & 3 & $5572^1$ & $4.5$ & $+0.15^2$ \\
HD20618 & G6 IV & 2 & $5600$ & $4.5$ & $-$ \\
HD101501 & G8 V & 2 & $5408^1$ & $4.5$ & $+0.03^2$ \\
HD185144 & K0 V & 2 & $5212^1$ & $4.5$ & $-0.29^1$ \\
HD22049 & K2 V & 2 & $5117^1$ & $4.6$ & $-0.14^2$ \\
HD39715 & rK3 V & 1 & $4850$ & $4.6$ & $+0.33^3$ \\
HD131977 & K4 V & 1 & $4700$ & $4.6$ & $+0.03^2$ \\
HD201902 & K7 V & 1 & $4100$ & $4.6$ & $-0.63^2$ \\
GL338 & M0 V & 1 & $3900$ & $4.6$ & $-$ \\
\hline
HD89025 & F0 III & $80$ & $7100$ & $3.4$ & $-$ \\
HD432 & F2 III & $75$ & $7278^1$ & $3.2$ & $+0.18^1$ \\
HD107950 & G6 III & $4$ & $5050$ & $2.8$ & $-0.16^2$ \\
HD197989 & K0 III & $2$ & $4800$ & $2.7$ & $-0.18^2$ \\
HD3627 & K3 III & $2$ & $4300$ & $2.0$ & $+0.04^2$ \\
\hline
\end{tabular}

\end{table*}

\subsection{Comparisons}

\begin{figure*}
\centering
\includegraphics[angle=-90,width=12cm]{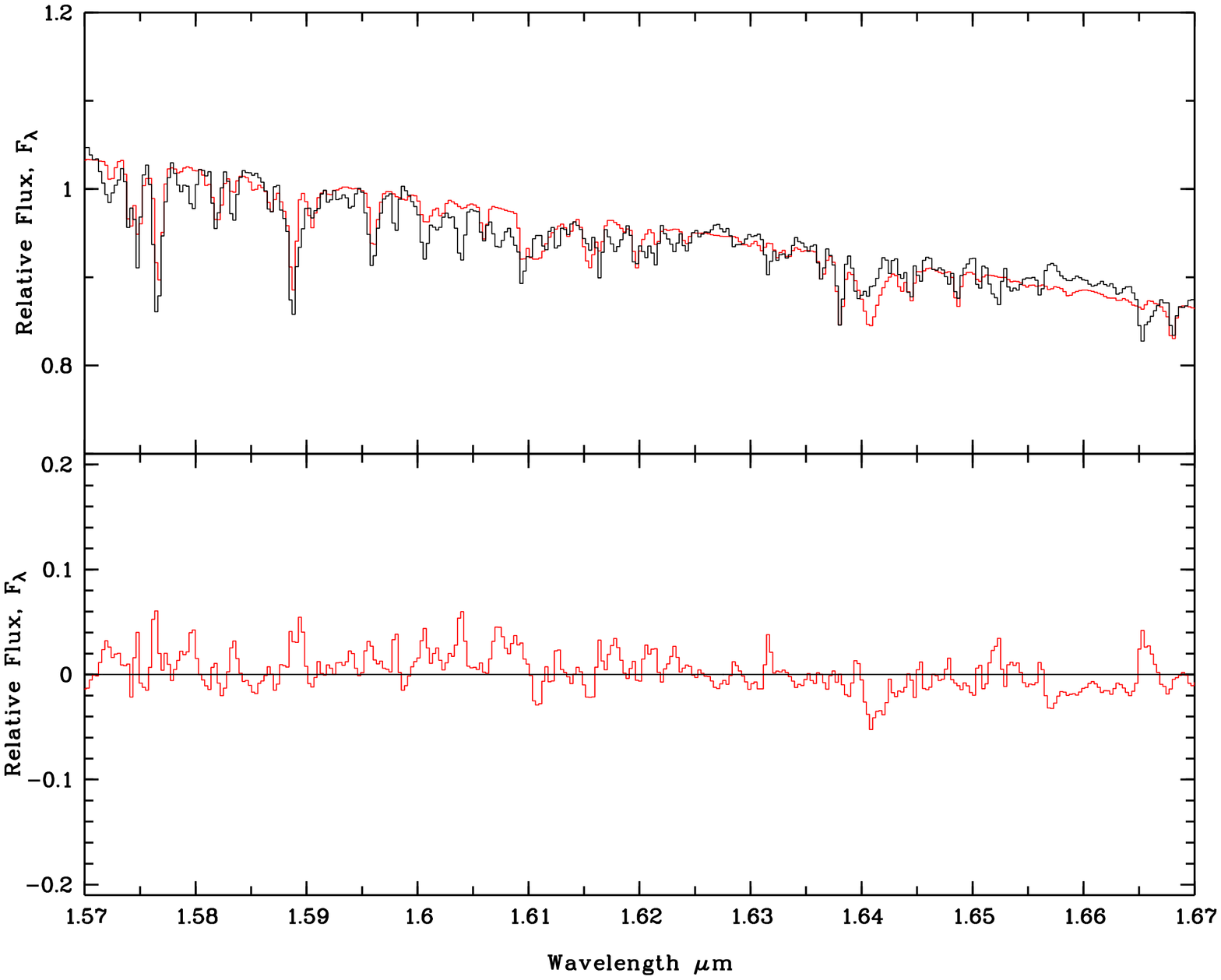}
\caption{The observed spectrum (in black) is from a F8.5V type star (HD 98231), the
         theoretical one (in grey) is computed with the following parameters:
		 T=5800K, log g=4.6, [M/H]=-0.3dex.}
\label{ir-f85v}
\end{figure*}

Around 1.6 $\mu$m hot stars are dominated by the Brackett lines, and a
good determination of the rotational velocity of these stars, which
broadens the lines, is very important to have the best possible match
between the computation and the observation. In our wavelength range,
the Brackett lines at 1.588, 1.611 and 1.641 $\mu$m are nearly the
only features of the observed spectra. They are well fitted by the
theoretical spectra.

When the temperature decreases, some atomic features appear and the
comparison between observed and computed spectra deteriorates.
Indeed, for the F6V, a quite large amount of metallic lines, visible
in the observed spectra, are not present or too weak in the computed
spectra and this trend continues with the F8.5V (Fig. \ref{ir-f85v}).
The continuum of these observed spectra is very well reproduced, but
this is not the case for the lines: most of the metallic lines are
computed too weak.

In addition, for the F8.5V star, the Brackett lines at 1.611 and 1.641
$\mu$m, are computed too strong. Indeed, the Brackett lines have
almost disappeared in the observed spectrum but are still strong in
the computation.

\begin{figure*}
\centering
\includegraphics[angle=-90,width=12cm]{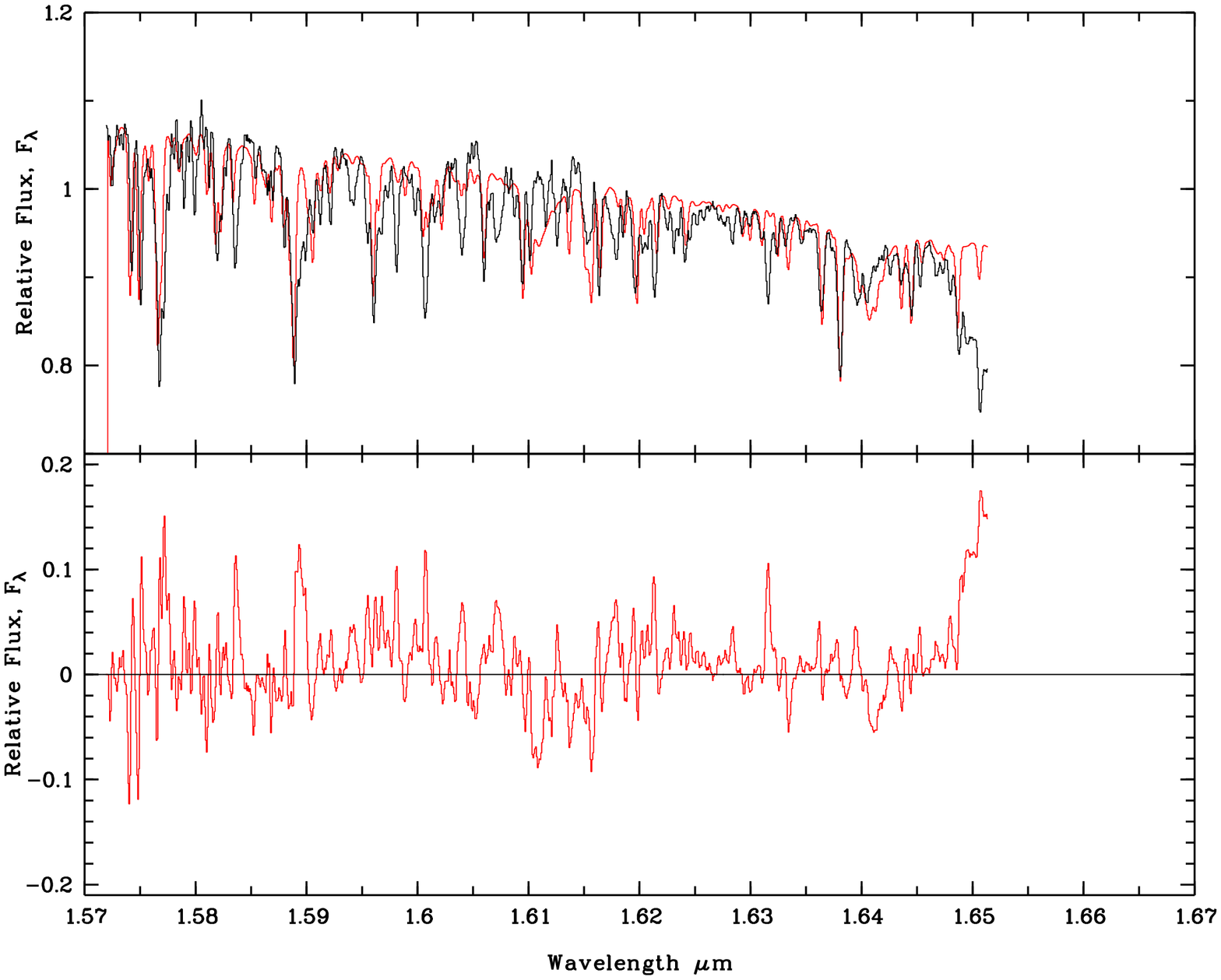}
\caption{The observed spectrum (in black) is from a G4V type star (HD 106116), the
         theoretical one (in grey) is computed with the following parameters:
		 T=5800K, log g=4.4, [M/H]=+0.3dex.}
\label{ir-g4v}
\end{figure*}

These Brackett lines are also present in all the theoretical G stars,
which is not always the case for observed stars, as seen in
Fig.~\ref{ir-g4v}. The behaviour of the computed Brackett lines
towards the temperature is shown on Fig.~\ref{ir-kgi}; the Brackett
lines are still present in theoretical spectra for temperatures as low
as 5200K.

\begin{figure*}
\centering
\includegraphics[angle=-90,width=12cm]{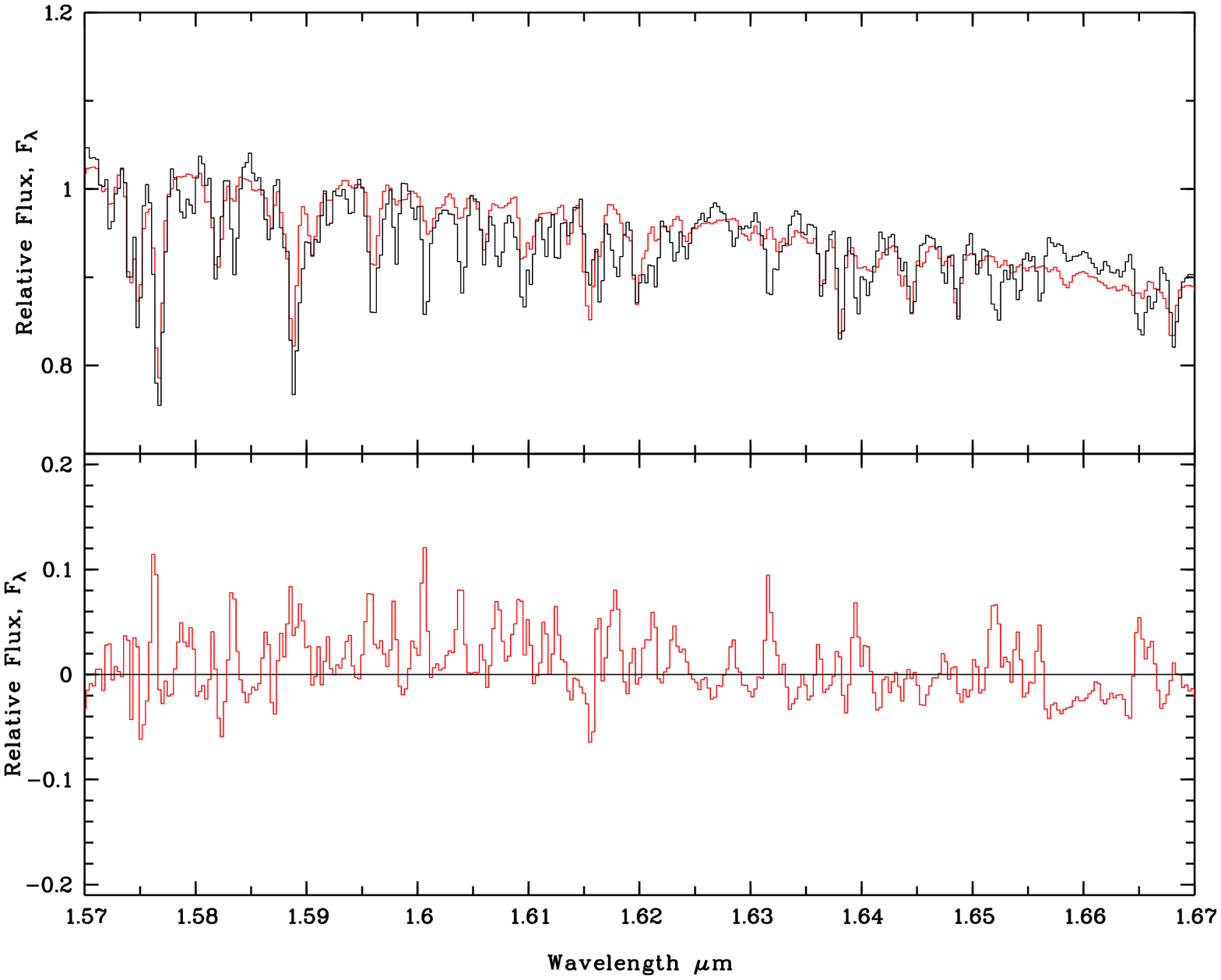}
\caption{The observed spectrum (in black) is from a K2V type star (HD 22049), the
         theoretical one (in grey) is computed with the following parameters: T=4800K,
         log g=4.6, [M/H]=-0.1dex.}
\label{ir-k2v}
\end{figure*}

\begin{figure*}
\centering
\includegraphics[angle=-90,width=12cm]{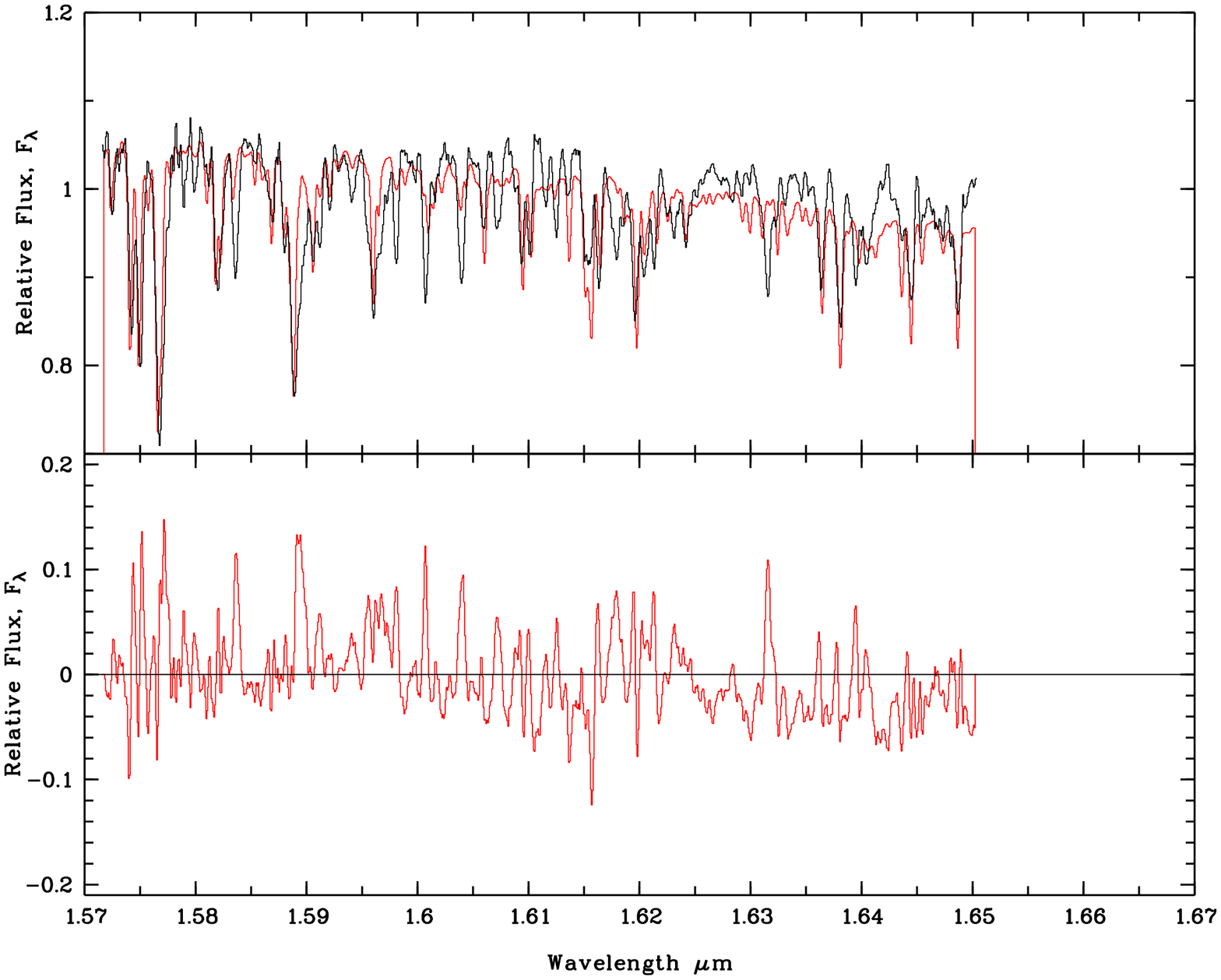}
\caption{The observed spectrum (in black) is from a K4V type star (HD 131977), the
         theoretical one (in grey) is computed with the following parameters: T=4800K,
		 log g=4.6, solar metallicity.}
\label{ir-k4v}
\end{figure*}

\begin{figure*}
\centering
\includegraphics[angle=-90,width=12cm]{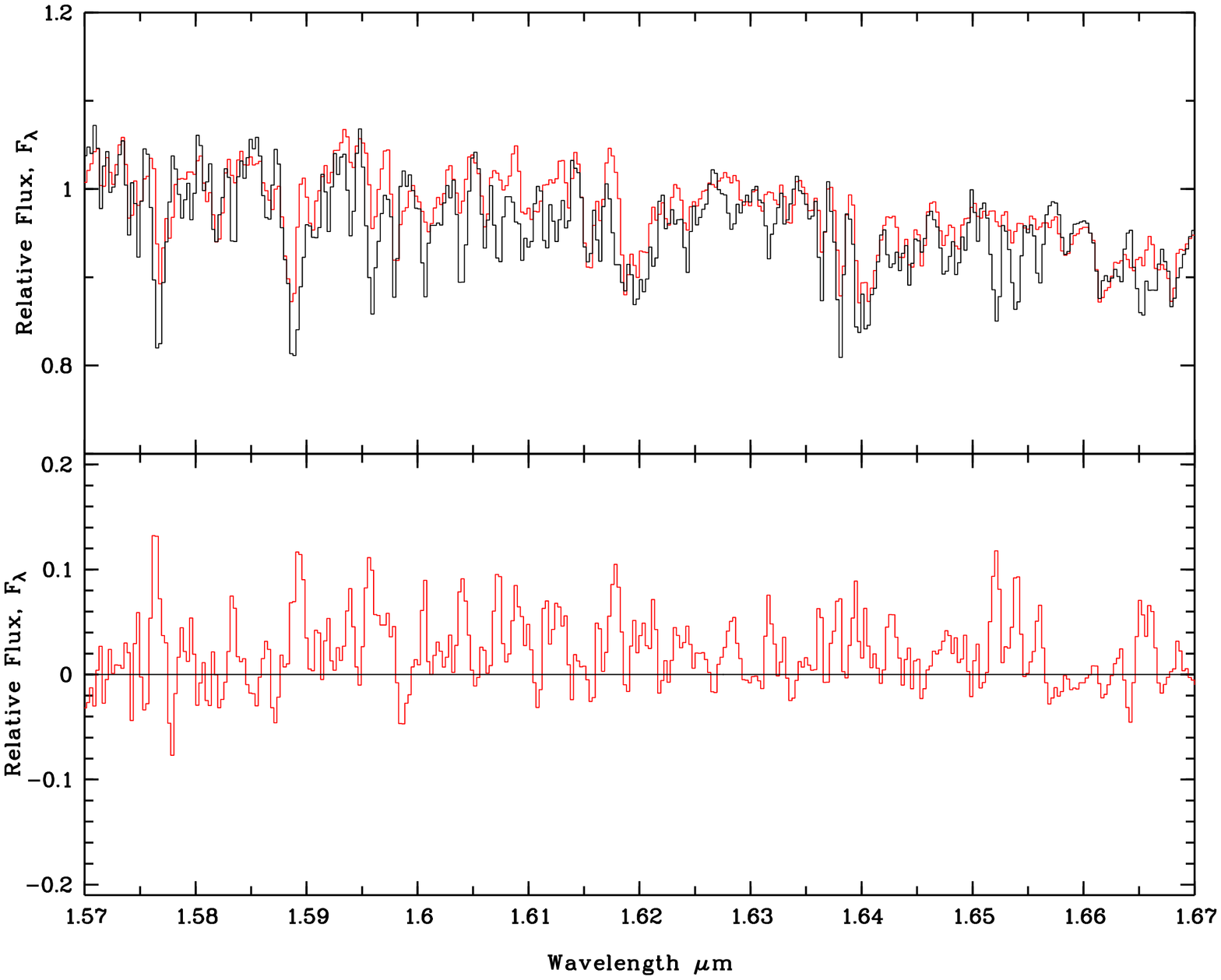}
\caption{The observed spectrum (in black) is from a K3III type star (HD 3627), the
         theoretical one (in grey) is computed with the following parameters: T=4400K,
		 log g=2.0, solar metallicity.}
\label{ir-k3iii}
\end{figure*}

Then, when the temperature decreases, as expected, the Brackett lines
are fainter and atomic lines (FeI, SiI, MgI and CaI) become stronger
and stronger for the observed spectra, as well as the theoretical
spectra.  But the agreement between the computed and the observed
stars becomes worse.  From Fig.~\ref{ir-g4v} to \ref{ir-k3iii}, the
residuals between observed and theoretical spectra show that several
absorption lines are missing in the theoretical stars, iron lines for
the most part. The model atmosphere is not the reason because the
continuum shape is very good and several lines match perfectly, but
the line list needs to be improved.

In the infrared range, the lack of several metallic and molecular
lines causes the discrepancies, with enhanced differences at low
temperature due to the greatest strength of the lines for the coolest
star (Fig. \ref{ir-k3iii}).  Fig.~\ref{ir-k2v} and \ref{ir-k4v}
present two similar stars (K2V and K4V, respectively), the first one
is from Meyer and the second one is a VLT observation at higher
resolution. Both comparisons show this lack of absorption lines, with
more details visible for Fig. \ref{ir-k4v}.

\begin{figure*}
\centering
\includegraphics[angle=-90,width=12cm]{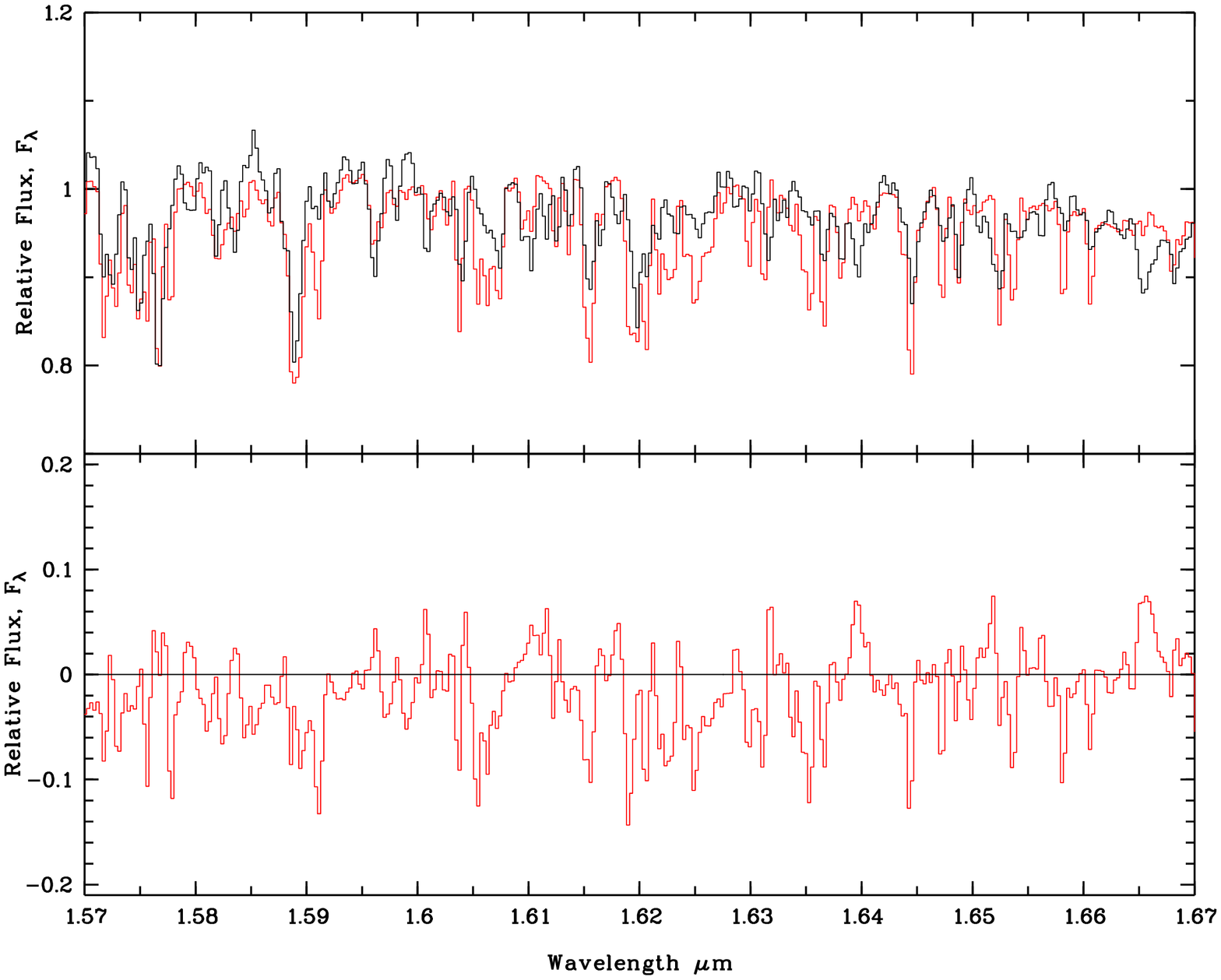}
\caption{The observed spectrum (in black) is from a M0V type star (GL 338), the
         theoretical one (in grey) is computed with the following parameters: T=4000K,
		 log g=4.6, solar metallicity.}
\label{ir-m0v}
\end{figure*}

The comparison for the coolest dwarf star of this sample, a M0V, is
not so bad for such a low temperature. The continuum is good in spite
of the limitations of the model (Fig.~\ref{ir-m0v}).  We notice,
however, that at the contrary of the previous spectra, the absorption
lines are computed too strong for the theoretical spectrum, as seen
thanks to the residual.  This is probably due to the limit of validity
of the model atmospheres, as already seen in the visible range.

\bigskip

In order to investigate further which lines are missing in the
computations, we have compared the high-resolution spectrum of the
well-known K1III star Arcturus (Hinkle et al. 1995) to a theoretical
spectrum computed with the parameters of this star ($T_\mathrm{eff}$ =
4400 K, log g = 2.0, [M/H]=-0.2, $v \sin i$ = 3.5 km/s) and point out
the discrepancies: Fig. \ref{ir_arct_zoom} shows a detail of this
comparison, and Table \ref{arct_lines} lists the missing lines in the
whole range. 
In addition to the lines quoted in Table \ref{arct_lines}, several
other features are computed too weak, in particular OH and CO
molecular bands, certainly due to an inaccurate determination of the
oscillator strengths, as discussed in Lyubchik et al. (\cite{Lyu04}).

\begin{figure*}
\centering
\includegraphics[angle=-90,width=12cm]{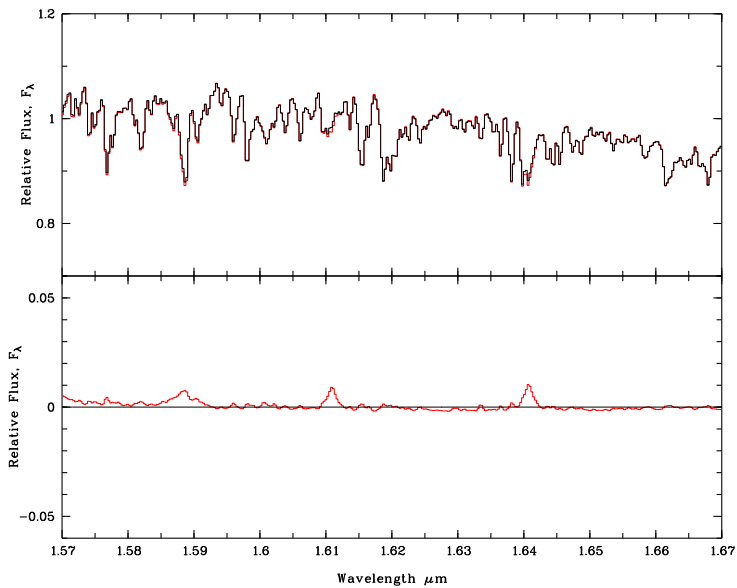}
\caption{Comparison between two theoretical models (ATLAS9 and NeMo)
  with the following parameters: T=4400K, log g=2.0, solar
  metallicity. Note that the scale for the residual is enhanced
  compared to all other figures.}
\label{nemo_kur}
\end{figure*}

This study, based on the NeMo grids of atmospheres, remains valid for the entire
familly of the ATLAS models. Indeed, as shown in Fig.~\ref{nemo_kur}, two theoretical
spectra computed for the same physical parameters, with the NeMo grid and the ATLAS9
models with the overshooting prescription (Kurucz \cite{Kur93}, \cite{Kur98}; Castelli
et al.~\cite{Cas97}, respectively), are very similar at our spectral resolution. The
discrepancies between the two different theoretical spectra are very faint compared
to the discrepancies between the models and the observed spectra.

The same ATLAS9 models, but without overshooting (NOVER models, Castelli et
al.~\cite{Cas97}), present even less differences with the NeMo spectra, in particular
the slight discrepancy found for the Brackett lines disappear. They are more sensitive
than other lines to the fact that the Kurucz overshooting prescription changes the
temperatures at Rosseland optical depths of 0.1 to 0.5.

\begin{figure*}
\centering
\includegraphics[angle=-90,width=12cm]{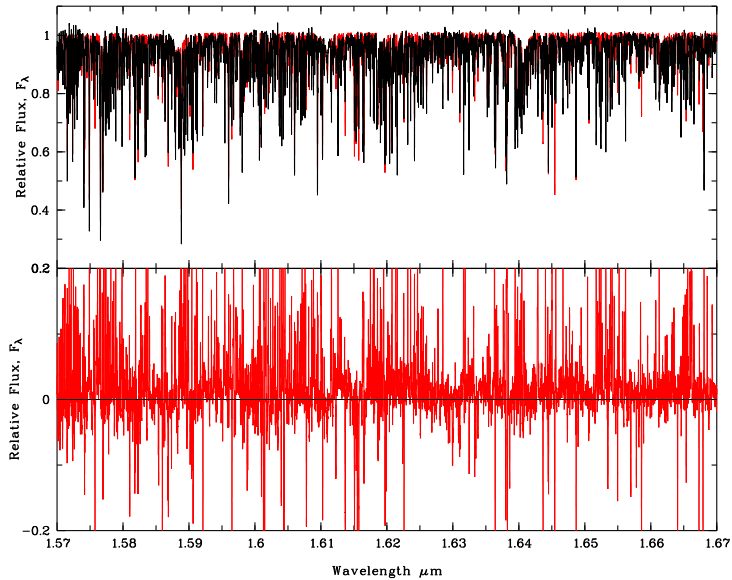}
\caption{Comparison between a high resolution observation
($\frac{\lambda}{\Delta\lambda} \simeq 100000$) of Arcturus (in
  black) and the corresponding theoretical star (in grey). The flux is
  given with continuum normalized to 1 in order to better show the
  missing lines.}
\label{ir_arct}
\end{figure*}

\begin{figure*}
\centering
\includegraphics[angle=-90,width=12cm]{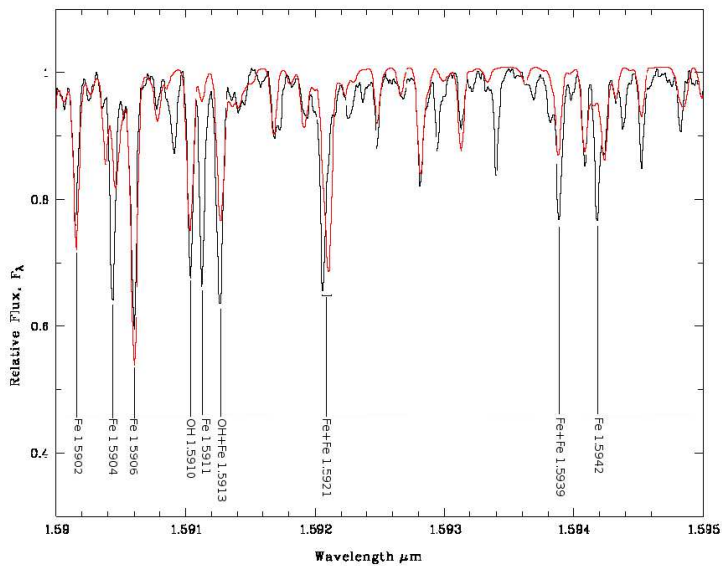}
\caption{The same as Fig. \ref{ir_arct} but zoomed in into a limited
         wavelength range.}
\label{ir_arct_zoom}
\end{figure*}

\begin{table}
\caption{List of the missing lines, according to the comparison between Arcturus and
          a computed spectrum. The Ni line is not missing as such but shifted by
		  2~\AA\ in the atomic database.}
\label{arct_lines}
\begin{tabular}{|c|c|c|c|c|}
\hline
wavelength ($\mu$m)& element & & wavelength ($\mu$m)& element \\
\hline

1.5764 & Fe & & 1.6208 & Fe \\
1.5893 & Fe & & 1.6214 & Fe \\
1.5895 & Fe & & 1.6231 & Fe \\
1.5913 & Fe & & 1.6285 & Fe \\
1.5939 & Fe & & 1.6316 & Fe \\
1.5954 & Fe & & 1.6319 & Fe \\
1.5968 & Fe & & 1.6362 & Ni \\
1.6007 & Fe & & 1.6394 & Fe \\
1.6008 & Fe & & 1.6440 & Fe \\
1.6041 & Fe & & 1.6450 & OH \\
1.6071 & Fe & & 1.6517 & Fe \\
1.6076 & Fe & & 1.6524 & Fe \\
1.6088 & Fe & & 1.6532 & Fe \\
1.6116 & Fe & & 1.6569 & Fe \\
1.6126 & Fe & & & \\
1.6175 & Fe & & & \\
1.6195 & Fe & & & \\
\hline
\end{tabular}
\end{table}

\section{Conclusions}  \label{conclusions}

In spite of some discrepancies, the comparisons between observed and
theoretical spectra in the visible range suggests a reasonable
agreement, even at the limits of the parameter range of NeMo.
However, we have to be careful when we compute a model near the lower
limit in temperature.

The spectra modelized with NeMo can be used to build a theoretical
spectral library for A to K dwarf and giant stars in the visible
range, but this is not the case for the near-infrared range.

Indeed, in the range 1.57 to 1.67 $\mu$m, the spectra computed do not reproduce
very well the observations. Albeit the good agreement for the overall flux
distribution shape, we can see that there are many differences for the line
features when focusing on details of the spectra. The strength of the infrared
absorption lines is usually underestimated in calculations, and some lines are
simply missing (Fe, OH and CO lines are the most problematic ones). As pointed
out by Decin et al.\ (\cite{Dec03}) for the MARCS6 models and
Lyubchik et al.\ (\cite{Lyu04}) for NextGen models of ultracool dwarfs and
a Kurucz model for Arcturus, it was not possible to generate synthetic spectra
which can reproduce observed spectra in the infrared with the line lists that
have been used in constructing the model atmospheres, even at a medium
resolution. In particular, the oscillator strengths are still not known
sufficiently well.

We have also performed a comparison of spectra that include the lines
from iron peak elements with predicted energy levels, as published by
Kurucz (\cite{Kur98}), with the observations of Arcturus discussed
above.  As to be expected, such spectra contain more lines, but the
inaccuracy of their energy levels frequently places them at the wrong
wavelengths and the overall match hardly improves (the total flux
distribution over all wavelength ranges, particularly in the
ultraviolet, is closer to observations when including this set of
lines, but for the limited wavelength range around the H band the
effects are small and sufficiently compensated when setting the zero
point of the flux distribution). For the case of Arcturus we also
performed a comparison with spectra computed with PHOENIX in LTE (P.\
Hauschildt, priv.\ comm.\ 2005). The resulting spectra for the 1.57 to
1.67~$\mu$m range were found to be rather similar to those from
NeMo/VALD/SynthV when including the predicted level lines.  One
important reason for this is certainly the fact that the atomic line
lists for PHOENIX are essentially those of Kurucz (\cite{Kur98}).
Because the flux distribution of the PHOENIX spectra is similar to the
observations as well, at least for the K giants the detailed choice of
the model atmosphere code appears to be clearly less important than
the choice of atomic line lists (note that PHOENIX uses its own
collection of molecular line lists, different from the one we have
used here). Considering the uniformity of the deterioration of the
match of spectra in the 1.57 to 1.67~$\mu$m range when looking at the
sequence from F to K stars we conclude that the insufficient line
lists, and in particular lists of atomic lines, are the main obstacle
for a more satisfactory match of observed spectra of these groups of
stars. The modelizing in the infrared range needs some further
improvements, in particular for the absorption lines database, before
to build a theoretical spectral library which can be used with high
benefit instead of an observed star library.

The lack of M stars in spectral library would be very much prejudicial
to the study of stellar populations as the variations of their strong
atomic lines and molecular bands along their evolution from dwarf to
supergiant to giant provides very good age discriminators (from 10$^6$
to 10$^{10}$yrs). M stars peak in a wavelength range which is not much
absorbed even in heavily reddened region, as young stellar clusters,
making them easily detectable. Moreover they are known to be very
important contributors to the stellar populations of galaxies as well
for the mass as for the luminosity following the age of the
population. All this make a good theoretical library of M stars very
critical in order to extend incomplete observed library. 

The prospects of constructing such a library from the upcoming generation
of model atmospheres (MARCS, PHOENIX, perhaps future versions of ATLAS)
are indeed improving because of the enormous efforts spent in extending
the molecular line data and equation of state. To match
the spectra of the hot end of M stars in the H band will nevertheless
require more complete atomic data, although this is less crucial as for
K stars. Efforts along this direction are currently made.

\begin{acknowledgements}
FK gratefully acknowledges the hospitality of the Observatoire de Paris-Meudon
during his stays as an invited visitor. 
JF also thanks the nice people of the AMS group at Vienna Observatory for 
their hospitality and help during part of this work. 
VT acknowledges the Austrian Fonds zur F\"orderung der wissenschaftlichen
Forschung FwF (P17580) and by the BM:BWK (project COROT).
This research has made use of the model atmosphere grid NeMo, provided by
the Department of Astronomy of the University of Vienna, Austria, and
funded by the Austrian FwF (P14984).
We are thankful to Peter Hauschildt who has computed comparison spectra for
Arcturus for us which helped to demonstrate the importance of the line lists
used relative to the particular choice of model atmosphere codes.
\end{acknowledgements}


\begin{thebibliography}{}

\bibitem[1995]{All95}
Allard, F., \& Hauschildt, P.H. 1995, ApJ, 445, 433
\bibitem[1989]{And89}
Anders, E., \& Grevesse, N. 1989, Geochim. Cosmochim. Acta, 53, 197
\bibitem[2003]{Bag03}
Bagnulo, S., Jehin, E., Ledoux, C., Cabanac, R., Melo, C., Gilmozzi, R., The ESO Paranal Science Operations Team 2003, The Messenger, 114, 10
\bibitem[2002]{Bar02}
Baraffe, I., Chabrier, G., Allard, F., \& Hauschildt, P.H. 2002, A\&A, 382, 563
\bibitem[1990]{Bar90}
Barbuy, B., \& Grenon, M. 1990, In: ESO/CTIO Workshop on Bulges of Galaxies. (A92-18101 05-90),p.83
\bibitem[2004]{Ber04}
Bertone, E., Buzzoni, A., Chavez, M., Rodriguez-Merino, L.H. 2004, AJ, 128, 829
\bibitem[2000]{Boi00}
Boisson, C., Joly, M., Moultaka, J., Pelat, D., \& Serote Roos, M. 2000, A\&A 357, 850
\bibitem[2002]{Boi02}
Boisson, C., Coup\'e, S., Cuby, J. G., Joly, M., Ward, M. J. 2002, A\&A, 396, 489
\bibitem[2003]{Bru03}
Bruzual, G., \& Charlot, S. 2003, MNRAS, 344, 1000
\bibitem[1996]{Can96}
Canuto, V.M., Goldman, I., \& Mazzitelli, I. 1996, ApJ, 473, 550
\bibitem[1991]{Can91}
Canuto, V.M., \& Mazzitelli, I. 1991, ApJ, 370, 295
\bibitem[1992]{Can92}
Canuto, V.M., \& Mazzitelli, I. 1992, ApJ, 389, 724
\bibitem[1997]{Cas97}
Castelli, F., Gratton, R., \& Kurucz, R.L. 1997, A\&A, 318, 841 (erratum: 1997, A\&A, 324, 432)
\bibitem[2003]{Cas03}
Castelli, F., \& Kurucz, R.L. 2003, in Modelling of Stellar Atmospheres, IAU Symposium vol. 210, eds. N.E. Piskunov, W.W.Weiss \& D.F. Gray,
p. A20C
\bibitem[2001]{Cay01}
Cayrel de Strobel G., Soubiran C., Ralite N. 2001, A\&A, 373, 159
\bibitem[2005]{Cid05}
Cid Fernandes, R., Mateus, A., Sodr\'e, L., Stasi\'nska, G., Gomes, J.M. 2005, MNRAS, 358, 363
\bibitem[1996]{Dal96}
Dallier, R., Boisson, C., \& Joly, M. 1996, A\&A SS, 116, 239
\bibitem[2003]{Dec03}
Decin, L., Vandenbussche, B., Waelkens, C., Eriksson, K., Gustafsson, B., Plez, B., Sauval, A.J., \& Hinkel, K. 2003, A\&A, 400, 679
\bibitem[2000]{Dec00}
Decin, L., Waelkens, C., Eriksson, K., Gustafsson, B., Plez, B., Sauval, A.J., Van Assche, W. \& Vandenbussche, B. 2000, A\&A, 364, 137
\bibitem[1993]{Fur93}
Fuhrmann, K., Axer, M., \& Gehren, T. 1993, A\&A, 271, 451
\bibitem[1992]{Gra92}
Gray, D.F. 1992, The observation and analysis of stellar photospheres (Cambridge University Press)
\bibitem[1975]{Gus75}
Gustafsson, B., Bell, R.A., Eriksson, K., Nordlund, {\AA} 1975, A\&A, 42, 407
\bibitem[1999]{Hau99}
Hauschildt, P.H., Allard, F., Ferguson, J., Baron, E., \& Alexander, D.R. 1999, ApJ, 525, 871
\bibitem[2002]{Hei02}
Heiter, U., Kupka, F., van't Veer-Menneret, C., Barban, C., Weiss, W.W., Goupil, M.-J., Schmidt, W., Katz, D., \& Garrido, R. 2002, A\&A, 392, 619
\bibitem[1995]{Hin95}
Hinkle, K., Wallace, L., \& Livingston, W. 1995, PASP, 107, 1042
\bibitem[2001]{Hol01}
Holweger, H. 2001, In: SOHO/ACE Workshop "Solar and Galactic Composition", R.F. Wimmer-Schweingruber (eds.),
AIP Conference Series 598 (Springer, New York), p. 23
\bibitem[1988]{Hub88}
Hubeny, I. 1988, Computer Physics Comm., 52, 103
\bibitem[1995]{Hub95}
Hubeny, I., \& Lanz, T. 1995, ApJ, 439, 875
\bibitem[2004]{Iva04}
Ivanov, V.D., Rieke, M.J., Engelbracht, C.W., Alonso-Herrero, A., Rieke, G.H., Luhman, K.L. 2004, ApJS, 151, 387
\bibitem[1999]{Kup99}
Kupka, F., Piskunov, N.E., Ryabchikova, T.A., Stempels, H.C., \& Weiss, W.W. 1999, A\&AS, 138, 119
\bibitem[1992]{Kur92}
Kurucz, R.L. 1992, in The Stellar Population of Galaxies, IAU Symp. 149,
eds. Barbuy B., Renzini A., Kluwer, Dordrecht, p. 225
\bibitem[1993a]{Kur93}
Kurucz, R.L. 1993a, ATLAS9 Stellar atmospheres programs and 2km/s grid, CD-ROM 13, SAO
\bibitem[1993b]{Kur93b}
Kurucz, R.L. 1993b, Atomic data for molecules, CD-ROM 15, SAO
\bibitem[1998]{Kur98}
Kurucz, R.L. 1998, http://kurucz.harvard.edu/, http://cfaku5.cfa.harvard.edu/
\bibitem[1999]{Kur99}
Kurucz, R.L. 1999, Atomic data for TiO and H2O, CD-ROMs 24, 25 and 26, SAO
\bibitem[2003]{Lan03}
Lanz, T., \& Hubeny, I. 2003, ApJS, 146, 417
\bibitem[2004]{LeB04}
Le Borgne, D., Rocca-Volmerange, B., Prugniel, P., Lan\c{c}on, A., Fioc, M., Soubiran, C. 2004,
\bibitem[2003]{LeB03}
Le Borgne, J.-F., Bruzual, G., Pell\'o, R., Lan\c{c}on, A., Rocca-Volmerange, B., Sanahuja, B., Schaerer, D., Soubiran, C., V\'ilchez-G\'omez, R. 2003, A\&A, 402, 433
\bibitem[1999]{Lei99}
Leitherer, C., Schaerer, D., Goldader, J.D., Gonz\'alez Delgado, R.M., Robert, C., Kune, D.F., de Mello, D.F., Devost, D., Heckman, T.M. 1999, ApJS, 123, 3
\bibitem[2004]{Lyu04}
Lyubchik, Y., Jones, H.R.A., Pavlenko, Y.V., Viti, S., Pickering, J.C., \& Blackwell-Whitehead, R. 2004, A\&A, 416, 655
\bibitem[2005]{Mar05}
Martins, L.P., Gonz\'alez Delgado, R.M., Leitherer, C., Cervi\~no, M., \& Hauschildt, P. 2005, MNRAS, 358, 49
\bibitem[1998]{Mey98}
Meyer, M.R., Edwards, S., Hinkle, K.H. \& Strom, S.E. 1998, ApJ, 508, 397
\bibitem[2000]{Mou00}
Moultaka, J., \& Pelat, D. 2000, MNRAS 314, 409
\bibitem[2005]{Mun05}
Munari, U., Sordo, R., Castelli, F., \& Zwitter, T. 2005, A\&A, \textit{in press}, astro-ph/0502047
\bibitem[2004]{Mur04}
Murphy, T., \& Meiksin, A. 2004, MNRAS, 351, 1430
\bibitem[2003]{NeMo}
NeMo website, 2003, http://ams.astro.univie.ac.at/nemo/
\bibitem[2004]{Nend04}
Nendwich, J., Heiter U., Kupka F., Nesvacil N., Weiss W.W. 2004,
Comm. Asteroseism., 144, 43
\bibitem[2004]{Nor04}
Nordstr\"{o}m, B., Mayor, M., Andersen, J., Holmberg, J., Pont, F., J\o rgensen, B. R., Olsen, E. H., Udry, S., \& Mowlavi, N. 2004, A\&A, 418, 989
\bibitem[1997]{Pel97}
Pelat D. 1997, MNRAS 284, 365
\bibitem[1992]{Ple92}
Plez, B., Brett, J.M., \& Nordlund, {\AA}. 1992, A\&A, 256, 551
\bibitem[2005]{Ram05}
Ram\'irez, I., \& Mel\'endez, J. 2005, ApJ, 626, 446
\bibitem[1982]{Sch82}
Schmidt-Kaler, Th. 1982, In: Schaifers K., Voight H.H. (eds)Landolt-B\"{o}rnstein; Stars and star clusters. Numerical data and
functional relationships in science and technology. Group IV, Vol. 2b
\bibitem[1996]{Ser96}
Serote Roos, M., Boisson, C., \& Joly, M. 1996, A\&AS, 117, 93
\bibitem[1992]{Sil92}
Silva, D., \& Cornell, M. 1992, ApJS, 81, 865
\bibitem[1997]{Sto97}
Stothers, R.B., \& Chin, C. 1997, ApJ, 478, L103
\bibitem[2004]{Val04}
Valdes, F., Gupta, R., Rose, J.A., Singh, H.P., \& Bell, D.J. 2004, ApJSS, 152, 251
\bibitem[1996]{vVe96}
van't Veer-Menerret, C., \& M\'egessier, C. 1996, A\&A, 309, 879
\bibitem[2005]{Woo05}
Woolf, V.M., \& Wallerstein, G. 2005, MNRAS, 356, 963
\end{thebibliography}
\end{document}